\documentclass[aps,prd,superscriptaddress,amsmath,amssymb,twocolumn,fleqn,nofootinbib]{revtex4-1} %superscriptaddress mette le affiliation come apice
\usepackage{xcolor}
\usepackage{graphicx}% Include figure files
\usepackage{colordvi}
\usepackage{mathtools}% to define \abs
\usepackage{empheq}
\usepackage{bm}% 
\usepackage{pgf,tikz}
\usepackage{mathrsfs}
\usetikzlibrary{arrows}
\usepackage{caption}
\usepackage{subfig}
\usepackage{floatrow}
\usepackage{braket}
\usepackage{amssymb}
\usepackage{amsmath}
\usepackage[makeroom]{cancel}
\usepackage{hyperref}
\def\be{\begin{equation}}
\def\ee{\end{equation}}
\def\ba{\begin{eqnarray}}
\def\ea{\end{eqnarray}}

\newcommand{\dd}{\mathrm{d}}
\definecolor{wwffqq}{rgb}{0.4, 1.0, 0.0}
\definecolor{qqccqq}{rgb}{0.0, 0.8, 0.0}
\definecolor{ffttww}{rgb}{1.0, 0.2, 0.4}
\definecolor{ffqqtt}{rgb}{1.0, 0.0, 0.2}
\definecolor{ttffqq}{rgb}{0.2, 1.0, 0.0}
\definecolor{ududff}{rgb}{0.30196078431372547, 0.30196078431372547, 1.0}
\definecolor{xdxdff}{rgb}{0.49019607843137253, 0.49019607843137253, 1.0}
\begin{document}

\title{Energy-momentum tensor and helicity for gauge fields coupled to a pseudo-scalar inflaton}

\author{M. Ballardini}\email{mario.ballardini@gmail.com}
\affiliation{Dipartimento di Fisica e Astronomia, Alma Mater Studiorum
	Universit\`a di Bologna, \\
	Via Gobetti, 93/2, I-40129 Bologna, Italy}
\affiliation{INAF/OAS Bologna, via Gobetti 101, I-40129 Bologna, Italy}
\affiliation{INFN, Sezione di Bologna, Via Berti Pichat 6/2, I-40127 Bologna, Italy}
\affiliation{Department of Physics \& Astronomy, \\ 
	University of the Western Cape, Cape Town 7535, South Africa}
\author{M. Braglia}\email{matteo.braglia2@unibo.it}
\affiliation{Dipartimento di Fisica e Astronomia, Alma Mater Studiorum
Universit\`a di Bologna, \\
Via Gobetti, 93/2, I-40129 Bologna, Italy}
\affiliation{INAF/OAS Bologna, via Gobetti 101, I-40129 Bologna, Italy}
\affiliation{INFN, Sezione di Bologna, Via Berti Pichat 6/2, I-40127 Bologna, Italy}
\author{F. Finelli}\email{fabio.finelli@inaf.it}
\affiliation{INAF/OAS Bologna, via Gobetti 101, I-40129 Bologna, Italy}
\affiliation{INFN, Sezione di Bologna, Via Berti Pichat 6/2, I-40127 Bologna, Italy}
\author{G. Marozzi}\email{giovanni.marozzi@unipi.it}
\affiliation{Dipartimento di Fisica, Universit\`a di Pisa and INFN, Sezione di Pisa, Largo Pontecorvo 3, I-56127 Pisa, Italy}
\author{A. A. Starobinsky}\email{alstar@landau.ac.ru}
\affiliation{Landau Institute for Theoretical Physics, Moscow 119334, Russian Federation}
\affiliation{Kazan Federal University, Kazan 420008, Republic of Tatarstan, Russian Federation}

\date{\today}% It is always \today, today,
\begin{abstract}
We study the energy-momentum tensor and helicity of gauge fields coupled through $g \phi F \tilde{F}/4$ 
to a pseudo-scalar field $\phi$ driving inflation. Under the assumption of a constant time derivative 
of the background inflaton, we compute analitically divergent and finite terms of the energy density 
and helicity of gauge fields for any value of the coupling $g$.
We introduce a suitable adiabatic expansion for mode functions of physical states of the gauge 
fields which correctly reproduces ultraviolet divergences in average quantities and identify corresponding counterterms.
Our calculations shed light on the accuracy and the range of validity of approximated analytic 
estimates of the energy density and helicity terms previously existed in the literature in the 
strongly coupled regime only, i.e. for $g \dot \phi/(2H) \gg 1$.
We discuss the implications of our analytic calculations for the backreaction of quantum fluctuations 
onto the inflaton evolution.
\end{abstract}

\pacs{Valid PACS appear here}% PACS, the Physics and Astronomy
%Use showkeys class option if keyword
\maketitle
\section{Introduction}

Inflation driven by a real single scalar field (inflaton) slowly rolling on a smooth self-interaction 
potential represents the minimal class of models in General Relativity (GR) which are in agreement with observations.
Not only are the details of the fundamental nature of the inflaton and of its interaction with other 
fields needed to study the stage of reheating after inflation, but also they can be important for 
theoretical and phenomenological aspects of its evolution.

Axion inflation, and more generally, inflation driven by a pseudo-scalar field is the archetypal model 
to include parity violation during a nearly exponential expansion and it has a rich phenomenology. 
An interaction of the pseudo-scalar field with gauge fields of the type
\begin{equation}
\mathcal{L}_{\textup{int}} =- \frac{g\phi}{4} F^{\mu\nu}\tilde{F}_{\mu\nu} \,,
\label{eqn:coupling}
\end{equation} 
where $g$ is a coupling constant with a physical dimension of length, or inverse energy (we put 
$\hbar=c=1$), leads to decay of the pseudo-scalar field into gauge fields modifying its background 
dynamics \cite{Anber:2009ua} and to a wide range of potentially observable signatures including 
primordial magnetic fields 
\cite{Turner:1987bw,Garretson:1992vt,Finelli:2000sh,Anber:2006xt,Caprini:2014mja,Adshead:2016iae,Caprini:2017vnn,Chowdhury:2018mhj,Sobol:2019xls}, preheating at the end of inflation \cite{Finelli:2000sh,Adshead:2015pva,McDonough:2016xvu}     , baryogenesis and leptogenesis \cite{Jimenez:2017cdr,Domcke:2018eki,Domcke:2019mnd}, 
equilateral non-Gaussianites \cite{Barnaby:2010vf,Barnaby:2011vw,Linde:2012bt}, chiral gravitational 
waves in the range of direct detection by gravitational wave antennas \cite{Sorbo:2011rz,Barnaby:2011qe,Ferreira:2014zia,Peloso:2016gqs}, 
and primordial black holes (PBHs) 
\cite{Linde:2012bt,Bugaev:2013fya,Garcia-Bellido:2016dkw,Domcke:2016bkh,Domcke:2017fix}.

The decay of the inflaton into gauge fields due to the coupling in Eq.~\eqref{eqn:coupling}
is a standard problem of amplification of quantum fluctuation (gauge fields) in an external classical 
field (the inflaton). Applications of the textbook regularization techniques used for calculation of 
quantum effects in curved space-time \cite{Birrell:1982ix,GMM} 
have led to interesting novel results in the de Sitter \cite{Finelli:2004bm} and inflationary space-times 
\cite{Finelli:2001bn,Finelli:2003bp,Marozzi:2006ky}, also establishing a clear connection between the 
stochastic approach \cite{Starobinsky:1986fx,Starobinsky:1994bd} and field theory methods 
\cite{Finelli:2008zg,Finelli:2010sh}.

In this paper, we apply the technique of adiabatic regularization \cite{Zeldovich:1971mw,Parker:1974qw}\footnote{It was called 'n-wave regularization' in \cite{Zeldovich:1971mw}.} to 
the energy-density and helicity of gauge fields generated through the interaction in Eq.~\eqref{eqn:coupling}. 
The evolution equation of gauge fields admits analytical solutions under the assumption of a constant 
$\xi\equiv g \dot{\phi}/(2 H)$ \cite{Anber:2006xt}, where $H\equiv \frac{\dot a}{a}$ is the Hubble parameter during inflation, 
also considered constant in time. Here we solve in an analytical way for the averaged energy 
density and helicity of gauge fields for any value of $\xi$. 
Previously only approximate results valid in the strongly coupled regime $\xi \gg 1$ were obtained 
in the literature.
Our technique of computing integrals in the Fourier space is based on previous calculations of the 
Schwinger effect for a $U(1)$ gauge field in the de Sitter space-time
\cite{Frob:2014zka,Kobayashi:2014zza,Hayashinaka:2016qqn,Domcke:2019qmm},
but now it is applied to a novel problem in which the classical external field is the inflaton. 
More recent papers apply similar techniques to calculate backreaction of $SU(2)$ gauge fields 
\cite{Lozanov:2018kpk,Maleknejad:2018nxz} and fermions \cite{Adshead:2018oaa,Adshead:2019aac} on the de Sitter space-time.

Our paper is organized as follows. In Section~\ref{sec:setting} and \ref{Energy-momentum tensor} we review 
the basic equations and the averaged energy-momentum tensor and helicity of the gauge fields, respectively. 
In Section~\ref{Analytical calculation} we present analytical results for the bare averaged quantities, 
and we direct the interested reader to Appendix~\ref{Appendix5} for more detailed calculations. 
In Section~\ref{appendixADI} we outline the adiabatic regularization scheme used (see also \cite{Marozzi:2006ky}).
We also show that counterterms appearing in the adiabatic subtraction method can be naturally interpreted 
as coming from renormalization of self-interaction terms of the scalar field either existing in the bare 
Lagrangian density, or those which has to be added to it due to the 
non-renormalizability of the problem involved (that is clear from $g$ being dimensional).
We then describe the implications of our results to the homogeneous dynamics of inflation with the backreaction 
of one-loop quantum effects taken into account in Section~\ref{implicationsdissipative}, 
particularly focusing on the new regime of validity $|\xi|\lesssim 1$ and commenting on the differences from 
previous results existing in the literature. We then conclude in Section~\ref{Conclusions}.

\section{Setting of the problem}
\label{sec:setting}

The Lagrangian density describing a pseudo-scalar inflaton field $\phi$ coupled to a $U(1)$ gauge field is:
\be
\mathcal{L} = - \frac{1}{2} (\nabla\phi)^2 - V(\phi) 
- \frac{1}{4} (F^{\mu\nu})^2 - \frac{g\phi}{4} F^{\mu\nu}\tilde{F}_{\mu\nu} \,,
\label{eqn:pseudoscalar}
\ee
where $\tilde{F}^{\mu\nu}=\epsilon^{\mu\nu\alpha\beta}F_{\alpha\beta}/2=\epsilon^{\mu\nu\alpha\beta}
(\partial_\alpha A_\beta-\partial_\beta A_\alpha)/2$ and $\nabla$ is the metric covariant derivative. 
We consider the Friedmann-Lema$\rm\hat i$tre-Robertson-Walker (FLRW) metric $\dd s^2=-\dd t^2+a^2 \dd {\bf x}^2$, where $a(t)$ is the scale 
factor, and we write the coupling constant $g=\alpha/f$, where $f$ is the axion decay constant. 
We consider gauge fields to linear order in a background driven only by a non-zero time-dependent 
{\em vev} $\phi(t)$.

In this context, it is convenient to adopt the basis of circular polarization $\bm{\epsilon}_\pm$ transverse 
to the direction of propagation defined by the comoving momentum $\mathbf{k}$. In the Fourier space we then have:
\begin{align}
&\mathbf{k}\cdot \bm{\epsilon}_\pm = 0 \,,\\
&\mathbf{k}\times \bm{\epsilon}_\pm = \mp\imath|k|\bm{\epsilon}_\pm \, .
\end{align}
Expanding the second quantized gauge field in terms of creation and annihilation operators for each Fourier mode $\mathbf{k}$, we get: 
\be
\!\!\!\!\!\!\!\!\!\!\!\!\!\!
\mathbf{A} (t, \mathbf{x})
=\sum_{\lambda=\pm} \int \frac{\dd^3k}{(2\pi)^3} \left[ \bm{\epsilon}_\lambda(\mathbf{k})
A_\lambda(\tau,\mathbf{k})a_\lambda^{\mathbf{k}} e^{\imath \mathbf{k}\cdot\mathbf{x}} +H.c.\right] \,,
\ee
where the Fourier mode functions $A_\pm$ for the two circular polarizations satisfy the following equation of motion:
\be
\label{eqn:EoM1}
\frac{\dd^2}{\dd\,\tau^2}A_\pm(\tau,k)+\big( k^2 \bm{\mp} g k \phi' \big) A_\pm(\tau,k)=0\, .
\ee
Here the prime denotes the derivative with respect to the conformal time $\tau$ ($\dd \tau = \dd t /a$).

The above equation admits a simple analytic solution for a constant 
$\dot{\phi}$ $(\equiv d \phi/d t)$ in a nearly de Sitter stage during inflation. 
A constant time derivative for the inflaton evolution can be obtained for $V(\phi)=\Lambda^4(1-C|\phi|)$ with $|C\phi| \ll 1$,  
or for $V(\phi) \propto m^2 \phi^2$ \cite{St78,Linde:1983gd}.
Natural inflation with $V(\phi) = \Lambda^4 \left[ 1 \mp \cos (\phi/f) \right]$ can be approximated 
better and better by $m^2 \phi^2$ for $f \gg M_{pl}$ with $m=\Lambda^2/f$, which is the regime allowed by cosmic microwave background (CMB henceforth) anisotropy measurements 
\cite{Savage:2006tr,Akrami:2018odb}. 

We thus study the inflationary solution assuming a de Sitter expansion, 
i.e. $a(\tau)=-1/(H\tau)$, with $\tau < 0$, $H\simeq\textup{const}$   and $\dot{\phi}\simeq\textup{const}$. 
In such a case, we can write  $\phi' \simeq -\sqrt{2\epsilon_{\phi}}M_{pl}/\tau$  with $\epsilon_\phi = \dot{\phi}^2/(2 M_{\textup{pl}}^2 H^2)$ being one of the slow-roll parameters.
In this case, the equation of motion for the two circular polarization mode functions becomes 
\cite{Anber:2009ua}:
\be
\label{eqn:eqbeta}
\frac{\dd^2}{\dd\,\tau^2}A_\pm(\tau,k)+\Bigl(k^2\pm\frac{2 k \xi}{\tau}\Bigr)A_\pm(\tau,k)=0 \, ,
\ee
where $\xi \equiv g\dot{\phi}/(2H)$.
The above equation reduces to \cite[p.~538]{Abramovitz} with $L=0$ and admit a solution 
in terms of the regular and irregular Coulomb wave functions corresponding to the positive frequency for $-k \tau >0$:
\be
\label{eqn:solution}
A_\pm(\tau,k) = \frac{\left[G_0(\pm\xi,-k\tau)+\imath F_0(\pm\xi,-k\tau)\right]}{\sqrt{2k}} \,.
\ee
These can be rewritten in the subdomain $-k\tau>0$ in terms of the Whittaker W-functions:
\be
\label{eqn:aplusaminus}
A_\pm(\tau,k) = \frac{1}{\sqrt{2k}} e^{\pm\pi\xi/2} W_{\pm\imath\xi,\frac{1}{2}}(-2\imath k\tau) \,.
\ee
Note that the above solutions are symmetric under the change $A_+\to A_-$ and $\xi\to-\xi$ in Eq.~\eqref{eqn:eqbeta}.

%%%%%%%%%%%%%%%%%%%%%%%%%%%%%%%%%%%%%%%%%%%%%%%%%%%%%
\section{Energy-momentum tensor and helicity}
\label{Energy-momentum tensor}
%%%%%%%%%%%%%%%%%%%%%%%%%%%%%%%%%%%%%%%%%%%%%%%%%%%%%%

From the Lagrangian density in Eq.~\eqref{eqn:pseudoscalar}, it is easy to derive the 
metric energy-momentum tensor (EMT) for the gauge-fields:
\be
T_{\mu\nu}^{(F)} = F_{\rho\mu}F^\rho_{\,\,\,\,\nu}+g_{\mu\nu}\frac{\mathbf{E}^2-\mathbf{B}^2}{2}
\ee
with $\mathbf{E}$ and $\mathbf{B}$ being the associated electric and magnetic field.
We then obtain for the energy density and pressure the following expressions:
\begin{align}
\label{T00}
&T^{(F)}_{00} =  \frac{\mathbf{E}^2+\mathbf{B}^2}{2} \,, \\
\label{Tij}
&T^{(F)}_{ij} = -E_iE_j-B_iB_j+\delta_{ij}\frac{\mathbf{E}^2+\mathbf{B}^2}{2} \,.
\end{align}
Note that there is no terms in the EMT depending on the pseudo-scalar coupling. Their absence in the $T^{(F)}_{00}$ component follows from impossibility to construct a pseudo-scalar invariant under spatial rotations from $\mathbf{E}$ and $\mathbf{B}$. Then the fact that the EMT trace remains zero in the presence of the interaction of Eq.~\eqref{eqn:coupling}, due to the conformal invariance of the gauge field, leads to the absence of such terms in $T^{(F)}_{ij}$, too.

By using Eqs.~\eqref{T00} and \eqref{Tij}, the Friedmann equations take the form:
\begin{align}
\label{eqn:Friedmann}
&H^2 = \frac{1}{3 M_{pl}^2}\left[ \frac{\dot{\phi}^2}{2}+V(\phi)+\frac{\langle \mathbf{E}^2 + \mathbf{B}^2 \rangle}{2}
\right] \,, \\
\label{eqn:Einstein}
&\dot{H} = -\frac{1}{2M_{pl}^2} \left[\dot{\phi}^2+\frac{2}{3}\langle \mathbf{E}^2+\mathbf{B}^2 \rangle \right]\,.
\end{align}

Using the relations $\Vec{E}=-\Vec{A}'/a^2$ and $\Vec{B}=\Vec{\nabla}\times\Vec{A}'/a^2$, 
the averaged energy density is:
\begin{align}
        \label{eqn:integral2}
                \frac{\langle {\bf E}^2 + {\bf B}^2 \rangle}{2}
        & = \int\,\frac{\dd k}{(2\pi)^2\,a^4} I(k) \notag \\
        & = \int\,\frac{\dd k}{(2\pi)^2\,a^4} \,k^2\Bigl[|A_+'|^2 + |A_-'|^2  \notag\\
        & \quad \quad + k^2 (|A_+|^2 + |A_-|^2 ) \Bigr] \,.
\end{align}

The electric (magnetic) contribution is given by the first and second (third and fourth) term in 
the integrand. It is easy to see that this integral diverges for large momentum $k$. This is a common 
behavior for averaged quantities in quantum field theory (QFT henceforth) in curved background, or in external fields, and a renormalization procedure is needed 
to remove these ultraviolet (UV) divergences. In Section~\ref{appendixADI} we will use the adiabatic 
regularization method \cite{Zeldovich:1971mw,Parker:1974qw} for this purpose, and we will present counterterms needed to renormalize the bare constants in the 
Lagrangian in Section~\ref{counterterms}. In the present Section, we identify the UV divergent contributions in the integrands.

Expanding the integrand of Eq.~\eqref{eqn:integral2} for $-k \tau \gg 1$ we obtain quartic, quadratic 
and logarithmic UV-divergences:
\be \label{eqn:Idiv}
I_{\textup{div}}(k) \sim 2 k^3+\frac{\xi ^2 k}{\tau ^2}+\frac{3 \xi ^2 \left(-1+5 \xi ^2\right)}{4 \tau ^4 k}
+{\cal O} \left(\frac{1}{k}\right)^{3} \,.
\ee
It is interesting to note that the logarithmic divergence changes its sign when $\xi$ crosses 
$|\xi| = 1/\sqrt{5}$. On the other hand, expanding in the infrared (IR) limit ($-k \tau \ll 1$) 
the integrand of Eq.~\eqref{eqn:integral2} has no IR divergences.

The equation of motion for the inflation $\phi$ is affected by the backreaction of these gauge fields:
\be
\ddot \phi + 3 H \dot \phi + V_\phi = g \langle {\bf E} \cdot {\bf B} \rangle \,,
\label{KG}
\ee
where the helicity integral is given by:
\begin{align}
\label{eqn:integral}
\langle {\bf E} \cdot {\bf B} \rangle
= & - \int\,\frac{\dd k}{(2\pi )^2\,a^4} J(k) \notag\\
= & - \int\,\frac{\dd k\,k^3}{(2\pi )^2\,a^4}
\frac{\partial}{\partial \tau} \Bigl( |A_+|^2 - |A_-|^2 \Bigr) \,.
\end{align}

The integrand in Eq.~\eqref{eqn:integral} has a different divergent behavior compared to the energy 
density, since it has only quadratic and logarithmic divergences:
\be \label{eqn:Jdiv}
J_{\textup{div}}(k) \sim \frac{\xi k}{\tau^2}-\frac{3\xi(1-5\xi^2)}{2\tau^4 k} + {\cal O} \left(\frac{1}{k}\right)^{5/2} \,.
\ee
Also in this case the integrand in Eq.~\eqref{eqn:integral} does not have any IR pathology.
We point out that, even if we called it 'helicity integral', the above integral in Eq.~\eqref{eqn:integral} 
is actually the \emph{derivative} of what is usually called the helicity integral 
${\cal H} = \langle {\bf A} \cdot {\bf B} \rangle$ (which is also gauge-invariant for a coupling to a pseudo-scalar).

%%%%%%%%%%%%%%%%%%%%%%%%%%%%%%%%%%%%%%%%%
\section{Analytical calculation of divergent and finite terms}
\label{Analytical calculation}
%%%%%%%%%%%%%%%%%%%%%%%%%%%%%%%%%%%%%%%%%%

In order to find an analytical expression for the finite part, we note that the bare integrals in 
Eqs.~\eqref{eqn:integral2} and \eqref{eqn:integral} can be solved by using the expression of the mode 
functions $A_\pm$ given in Eq.~\eqref{eqn:aplusaminus}. 
We identify the divergences by imposing a UV physical cutoff $\Lambda$ in order to avoid time-dependent 
coupling constants at low energies: we therefore impose a comoving $k$-cutoff $k_{\textup{UV}}= \Lambda\, a$ 
\cite{Zeldovich:1971mw,Xue:2011hm} in the integrals in Eqs.~\eqref{eqn:integral2} and \eqref{eqn:integral}.
Note that Eq.~\eqref{eqn:eqbeta} can be solved analytically by Whittaker functions also in presence of 
a mass term \cite{Barnaby:2011qe}. %We restrict in this section to the results for the massless case, 
%but we also give the generalization to non-zero mass in Appendix~\ref{appendixmass}.
	
%%%%%%%%%%%%%%%%%%%%%%%%%%%%
\subsection{Energy Density}
\label{EnergyAnalytic}
%%%%%%%%%%%%%%%%%%%%%%%%%
We first compute the energy density (see Appendix~\ref{Appendix5} for details). With the help of the 
integral representation of the Whittaker functions and carefully choosing the integration contour we obtain 
for the energy density stored in the electric field:	
\begin{align}
        \label{electric}
       &\frac{\langle {\bf E}^2\rangle}{2}  =\frac{\Lambda^4}{16\pi^2}-\frac{\xi^2 H^2}{16\pi^2}\Lambda^2
        \notag\\
        &+\frac{ \xi ^2 \left(19-5 \xi ^2\right) H^4\log (2 \Lambda/H )}{32\pi^2}\notag\\
        &+\frac{H^4 \left(59 \xi ^6-470 \xi ^4-205 \xi ^2+36\right)}{384 \pi ^2 \left(\xi ^2+1\right)}\notag\\
        &-\frac{H^4   \left(30 \xi ^4-119 \xi^2+18\right) \sinh (2 \pi  \xi )}{384 \xi \pi ^3 }
        \notag\\
        &-\frac{i H^4 \xi ^2 \left(5 \xi ^2-19\right) \sinh (2 \pi  \xi ) \psi ^{(1)}(1-i \xi )}{128 \pi ^3 }\notag\\
        &+\frac{i H^4 \xi ^2 \left(5 \xi ^2-19\right) \sinh (2 \pi  \xi ) \psi^{(1)}(1+i \xi)}{128 \pi ^3 }\notag\\
               &+\frac{H^4 \xi ^2 \left(5 \xi ^2-19\right) \left(\psi(-1-i\xi)+\psi(-1+i\xi)\right)}{64 \pi ^2 } 
        \end{align}
while for the magnetic field we find
\begin{align}
	\label{magnetic}
	&\frac{\langle {\bf B}^2\rangle}{2}  =\frac{\Lambda^4}{16\pi^2}+\frac{3\xi^2 H^2}{16\pi^2}\Lambda^2 
	\notag\\
	&+\frac{5 \xi ^2 \left(7 \xi ^2-5\right) H^4\log (2 \Lambda/H )}{32\pi^2}\notag\\
	&-\frac{H^4 \left(533 \xi ^4-715 \xi ^2+36\right)}{384 \pi ^2 }\notag\\
	&+\frac{H^4 \left(210 \xi ^4-185 \xi ^2+18\right) \sinh (2 \pi  \xi )}{384 \pi ^3 \xi  }\notag\\
	&-\frac{5 H^4 \xi ^2 \left(7 \xi ^2-5\right) ( \psi(-i \xi )+\psi(i\xi))}{64 \pi ^2 }
	\notag\\
	&-\frac{5 i H^4 \xi ^2 \left(7 \xi ^2-5\right) \sinh (2 \pi  \xi ) \psi ^{(1)}(i \xi +1)}{128 \pi ^3 }\notag\\
	&+\frac{5 i H^4 \xi ^2 \left(7 \xi ^2-5\right) \sinh (2 \pi  \xi ) \psi^{(1)}(-i \xi +1)}{128 \pi ^3 }\,.
	\end{align}
Summing the two contributions, the total energy density becomes:	
	\begin{align}
	\label{eqn:energy}
	&\frac{\langle {\bf E}^2 + {\bf B}^2\rangle}{2}  =\frac{\Lambda^4}{8\pi^2}+\frac{\xi^2 H^2}{8\pi^2}\Lambda^2 
	\notag\\
	&+\frac{3 \xi ^2 \left(5 \xi ^2-1\right) H^4\log (2 \Lambda/H )}{16\pi^2}\notag\\
	&+\frac{\xi  \left(30 \xi ^2-11\right) \sinh (2 \pi  \xi )H^4}{64 \pi^3}\notag\\
	&+\frac{\xi ^2 \left(-79 \xi ^4+22 \xi ^2+29\right)H^4}{64 \pi^2 \left(\xi ^2+1\right) }
	\notag\\
	&+\frac{3 \imath \xi ^2 \left(5 \xi ^2-1\right) \sinh (2 \pi  \xi ) \psi ^{(1)}(1-\imath \xi )H^4}{64 \pi^3  }\notag\\
	&-\frac{3 \imath \xi ^2 \left(5 \xi ^2-1\right) \sinh (2 \pi  \xi ) \psi ^{(1)}(1+\imath \xi )H^4}{64 \pi^3  }\notag\\
	&-\frac{3 \xi ^2 \left(5 \xi ^2-1\right) [\psi (-\imath \xi -1)+\psi (\imath \xi -1)]H^4}{32\pi^2 },
	\end{align}
where $\psi$ is the Digamma function, $\psi^{(1)}(x)\equiv \dd \psi(x)/ \dd x$, 
$H_{x}\equiv\psi(x+1)+\gamma$ is the harmonic number of order $x$, and $\gamma$ is the 
Euler-Mascheroni constant. 
The finite terms in Eq.~\eqref{eqn:energy} have the corresponding asymptotic behavior:
	\begin{align}
	\label{energy_asymptotic}
	&  \frac{H^4}{64 \pi^2 }(19-16 \gamma)\xi^2 &\mathrm{when}\,\,\,\, &|\xi| \ll 1 \,, \\
	\label{energy_asymptotic2}
	& \frac{9 H^4 \sinh(2 \pi\xi)}{1120 \pi^3 \xi^3 }  &\mathrm{when}\,\,\,\, &|\xi| \gg 1\,.
	\end{align}

\begin{figure}[t]
	\centering
		\includegraphics[width=\columnwidth]{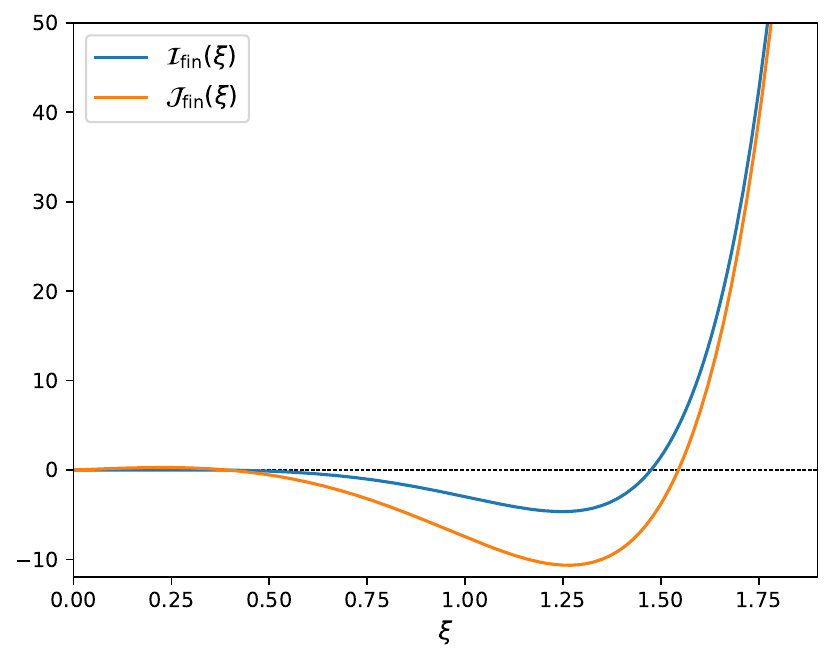}
		\caption{We plot respectively the quantities $\mathcal{I}_{fin}(\xi)$ (blue line) 
		and $\mathcal{J}_{fin}(\xi)$ (orange line) defined in Eqs.~\eqref{IFIN} and \eqref{JFIN}.
		\label{fig1}}
	\end{figure}

We now compare our results with those used in the literature which are based on the use of UV and IR cutoffs 
and an approximation of the integrand. More precisely Refs.~\cite{Anber:2006xt,Anber:2009ua} and subsequent 
works use the following approximation to estimate the integral:
	\begin{itemize}
		\item The integral has a physical UV cutoff at $-k\tau=2|\xi|$.
		\item Only the growing mode function $A_+$ in Eq.~\eqref{eqn:aplusaminus} is considered for $\xi > 0$ (the situation is reversed for $\xi < 0$) and it is approximated in this regime to:
		\begin{equation}
		A_+(\tau,k)\simeq\frac{1}{\sqrt{2 k}}\left(\frac{-k\tau}{2\xi}\right)^{1/4}e^{\pi\xi-\sqrt{-8\xi k \tau}} \,.
		\end{equation}
	\end{itemize}
Under this approximation Ref.~\cite{Anber:2006xt} obtains when $\xi\gg 1$:
	\begin{align}
	\frac{\langle {\bf E}^2 + {\bf B}^2 \rangle_\textup{AS}}{2}&
	\simeq 1.4 \cdot 10^{-4} 
\frac{H^4}{\xi^3} e^{2 \pi \xi} \,,
	\label{E2B2_literature}
	\end{align}
Our result in Eq.~\eqref{eqn:energy} is one of the main original results of this paper and is valid 
for any $\xi$.	
In Fig.~\ref{fig1}, we plot the terms of Eq.~\eqref{eqn:energy} which do not depend on the UV 
cut-off:
	\begin{align}
	\label{IFIN}
	&\mathcal{I}_{\textup{fin}}(\xi)\equiv \frac{\langle {\bf E}^2 + {\bf B}^2\rangle}{2} 
	-\Bigg[\frac{\Lambda^4}{8\pi^2}+\frac{\xi^2 H^2}{8\pi^2}\Lambda^2 \notag\\
	&+\frac{3 \xi ^2 \left(5 \xi ^2-1\right) H^4\log (2 \Lambda/H )}{16\pi^2}\Bigg] \,,
	\end{align}
in units of $(2\pi)^2/ H^4$ (for $\xi > 0$ for simplicity). These terms would correspond to the renomalized energy 
density obtained by a minimal subtraction scheme. 

\begin{figure}[t]
                \centering
               \includegraphics[width=\columnwidth]{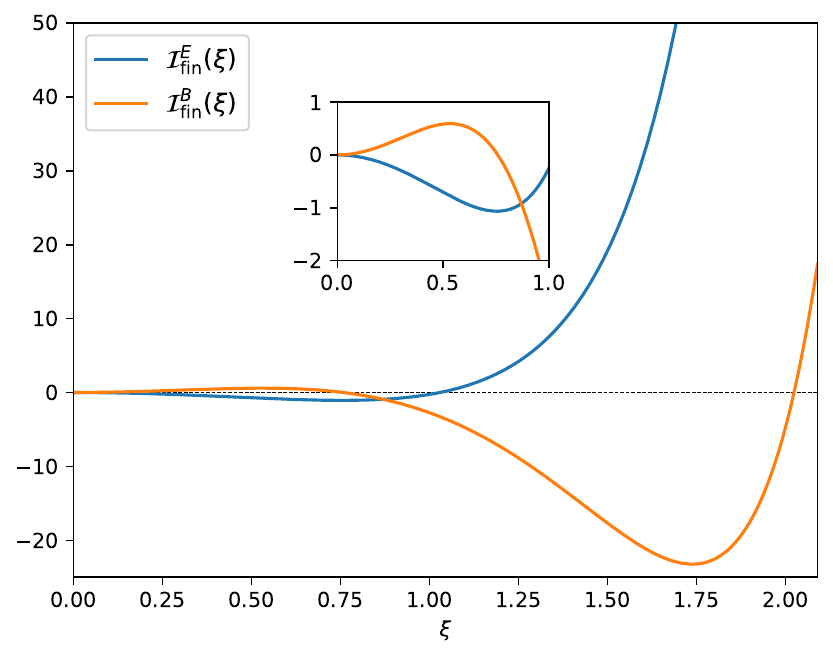}
                \caption{We plot in blue (orange) the electric (magnetic) contribution 
                to $\mathcal{I}_{fin}(\xi)$.    
\label{fig:ElMag}}
        \end{figure}

In Fig.~\ref{fig:ElMag}, we plot the electric and magnetic finite contributions to the energy density (by restricting to $\xi>0$, again). 
The electric contribution to the energy density is larger than the magnetic one for $\xi\gtrsim 0.75$, 
whereas they are comparable for $\xi\lesssim 0.75$.

\begin{figure}[t]
		\centering
	\includegraphics[width=\columnwidth]{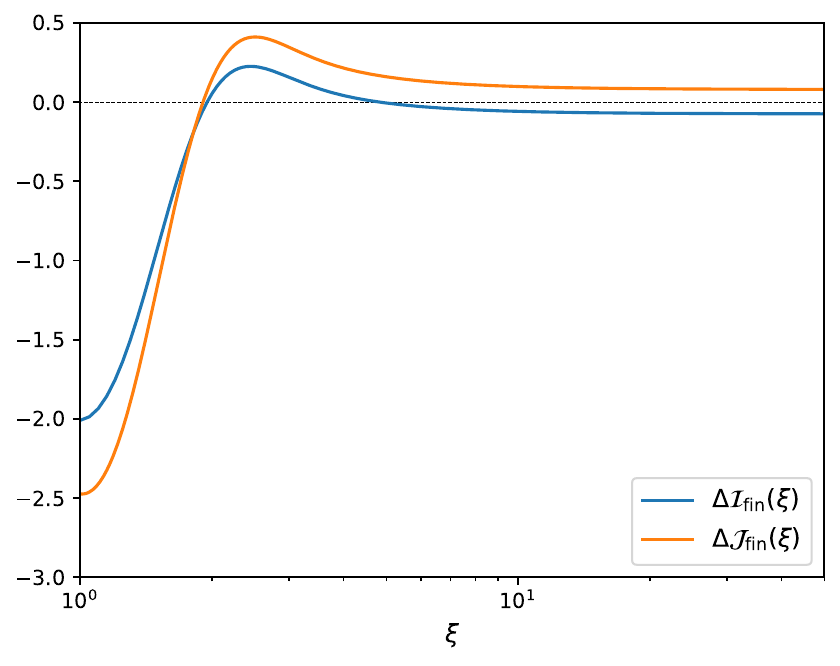}
		\caption{We plot respectively the quantities $\Delta\mathcal{I}_{fin}(\xi)$ (blue line) 
		and $\Delta\mathcal{J}_{fin}(\xi)$ (orange line) defined in Eqs.~\eqref{DIFIN} and \eqref{DJFIN}.
\label{fig2}}
	\end{figure}

In Fig.~\ref{fig2}, we plot the relative difference with Eq.~\eqref{E2B2_literature}
	\begin{equation}
	\label{DIFIN}
	\Delta\mathcal{I}_{\textup{fin}}(\xi)
\equiv\frac{2\mathcal{I}_{\textup{fin}}(\xi)-\langle {\bf E}^2 + {\bf B}^2 \rangle_\textup{AS}}{\langle {\bf E}^2 + {\bf B}^2 
\rangle_\textup{AS}}\,.
	\end{equation}
	
It can be seen from Eq.~\eqref{energy_asymptotic2} and Fig.~\ref{fig2} that the  behavior in the regime 
$|\xi|\gg1$ of our solution is similar to that of Eq.~\eqref{E2B2_literature}, which has thoroughly been studied in the literature. 
Nevertheless, Fig.~\ref{fig2} shows that there is a relative difference of approximately the order of 
10\% in the numerical coefficient that multiplies $\exp(2 \pi\xi)/\xi^3$, which can be ascribed to the 
assumptions described above. 

The main difference of our new result in Eq.~\eqref{eqn:energy} with respect to Eq.~\eqref{E2B2_literature} 
is in the regime of  $|\xi| \lesssim 10$, that has been studied for the first time in this paper. 
Eq.~\eqref{E2B2_literature} cannot be extrapolated to $|\xi|\ll1$, whereas our result shows that the 
finite part of the energy density is $\mathcal{O}(\xi^2)$ as shown in Eq.~\eqref{energy_asymptotic}. 
This difference can be understood by noting that the contributions from $A_+$ and $A_-$ become 
comparable in this regime of $\xi$ and neglecting $A_-$ is no longer a good approximation.

We end on noting that the finite contribution by a minimal subtraction scheme to the energy density, 
which is $\mathcal{O}(\xi^2)$ for $\xi \ll 1$, becomes negative for $0.8 \lesssim \xi \lesssim 1.5$, 
although its classical counterpart of Eq.~\eqref{eqn:energy} is positive definite. 
This is not totally surprising since it is known that in QFT in curved space-times
the renormalized terms of expectation values of classically defined positive terms can be negative 
\cite{Birrell:1982ix}.


%%%%%%%%%%%%%%%%%%%%%%%%%%%%%%%%%%%%	
\subsection{Helicity Integral}
%%%%%%%%%%%%%%%%%%%%%%%%%%%%%%%%
	
The helicity integral in Eq.~\eqref{eqn:integral} is only logaritmically and quadratically divergent 
because of the cancellation of the quartic divergence and does not exhibit any IR divergence. 
Note that only by considering both $A_+$ and $A_-$, quartic divergent terms in the UV regime cancel.
	
It is possible to derive an exact solution of Eq.~\eqref{eqn:integral} with a UV cutoff and we give 
the final result in the following, leaving the details for the interested reader in Appendix~\ref{Appendix5}.
The result for the helicity is:
	\begin{align}
	\label{edotb}
	&-\langle {\bf E} \cdot {\bf B} \rangle =\,\frac{ \xi  H^2}{8\pi^2 }\Lambda ^2+\frac{3 \xi  \left(5 \xi ^2-1\right)H^4 \log (2 \Lambda/H )}{8 \pi^2}\notag\\
	&+\frac{[6 \gamma  \xi  \left(5 \xi ^2-1\right)+(22 \xi -47 \xi ^3)]H^4}{16\pi^2}\notag\\
	&+\frac{\left(30 \xi ^2-11\right) \sinh (2 \pi  \xi )H^4}{32 \pi^3  }\notag\\
	&-\frac{3 \xi  \left(5 \xi ^2-1\right)   \left(H_{-\imath \xi }+H_{\imath \xi }\right)H^4}{16\pi^2 }\notag\\
	&+\imath\frac{3 \xi  \left(5 \xi ^2-1\right)  \sinh (2 \pi  \xi ) \psi ^{(1)}(1-\imath \xi )H^4}{32 \pi^3 }\notag\\
	&-\imath\frac{3 \xi  \left(5 \xi ^2-1\right)  \sinh (2 \pi  \xi ) \psi ^{(1)}(\imath \xi +1)H^4}{32 \pi^3 } \,.
	\end{align}
The finite terms have the corresponding asymptotic values:
	\begin{align}
	\label{back_reaction_asymptotic}
	& \frac{H^4}{16\pi^2}(11-6 \gamma)\xi &\mathrm{when}\,\,\,\, &\xi \ll 1  \,, \\
	\label{back_reaction_asymptotic2}
	& \frac{9 \sinh(2 \pi\xi)H^4}{560 \pi^3 \xi^4}  &\mathrm{when}\,\,\,\, &\xi\gg 1 \,.
	\end{align}
	
The result reported in the literature for the integral in Eq.~\eqref{eqn:integral} is derived under the same assumptions discuss in the context of Eq.~\eqref{E2B2_literature} and is given by 
\cite{Anber:2006xt}:
	\begin{align}
	-\langle {\bf E} \cdot {\bf B} \rangle_\textup{AS} \simeq 10^{-4}
	\frac{H^4}{\xi^4} e^{2 \pi \xi} \,.
	\label{EB_literature}
	\end{align}
Again, we define:
	\begin{align}
	\label{JFIN}
	&\mathcal{J}_{\textup{fin}}(\xi)\equiv -\langle {\bf E} \cdot {\bf B} \rangle\notag\\
	&-\left[\frac{ \xi  H^2}{8\pi^2 }\Lambda ^2+\frac{3 \xi  \left(5 \xi ^2-1\right)H^4 \log (2 \Lambda/H )}{8 \pi^2}\right] \,.
	\end{align}
and the relative differences between our solution and Eq.~\eqref{EB_literature}
	\begin{equation}
	\label{DJFIN}
	\Delta\mathcal{J}_{\textup{fin}}(\xi)\equiv-\frac{\mathcal{J}_{\textup{fin}}(\xi)+\langle {\bf E} \cdot {\bf B}\rangle_\textup{AS}}{\langle {\bf E} \cdot {\bf B}\rangle_\textup{AS}}\,,
	\end{equation}
which we plot in Fig.~\ref{fig1} and Fig.~\ref{fig2} respectively.
	
For $\xi \ll 1$ the back-reaction in Eq.~\eqref{back_reaction_asymptotic} is reminiscent of an extra dissipative term of the type $\Gamma_{\textup{dS}} \dot \phi$. It is interesting to note that the effective $\Gamma_{\textup{dS}}$ in this nearly de Sitter evolution is larger than the perturbative decay rate $\Gamma = g^2 m_\phi^3/(64 \pi)$ by a factor ${\cal O} (H^3/m_\phi^3)$.  
	
Analogously to the energy density case, Fig.~\ref{fig2} shows that in the regime of $\xi\gtrsim 10$ 
Eqs~.\eqref{back_reaction_asymptotic2} and \eqref{EB_literature} have a similar functional form, but 
still a $10\%$ difference. 
Our exact result can be more precisely used for $\xi\lesssim10$ and in particular in the $\xi\lesssim1$ 
regime. Note that the difference between the exact result and the result given in \cite{Anber:2006xt} is 
now larger than in the energy density case and that we have a linear dependence on $\xi$ for the helicity 
integral for $\xi\ll 1$, in a regime to which the standard result in the literature in 
Eq.~\eqref{EB_literature} cannot be extrapolated.

%%%%%%%%%%%%%%%%%%%%%%%%%%%%%%%%%%%%%%%%%%%%%%%%%%%%%%%%
\section{Adiabatic Expansion and Regularization}
\label{appendixADI}
%%%%%%%%%%%%%%%%%%%%%%%%%%%%%%%%%%%%%%%%%%%%%%%%%%

The adiabatic regularization method \cite{Zeldovich:1971mw,Parker:1974qw} relies on the
adiabatic, or Wentzel-Kramer-Brillouin (WKB henceforth), expansion of the mode functions $A_\pm$ solution of Eq.~\eqref{eqn:EoM1}. 
Following the standard adiabatic regularization procedure, we proceed by adding a mass term regulator $\mu$ to the evolution equations of the two different helicity states obtaining a modified version of Eq.~\eqref{eqn:EoM1}
\begin{equation}
\label{eqn:adievoeq}
\!\!\!\!\!\!\!\!\!\!\!\!\!\!\!
\frac{\dd^2}{\dd\,\tau^2}A_\pm^{\rm WKB}(k,\tau)+\Bigl(k^2\,\mp g k \phi'+\frac{\mu^2}{H^2 \tau^2}\Bigr)
A_\pm^{\rm WKB}(k,\tau)=0 \,,
\end{equation}

where the adiabatic mode function solution $A_\lambda^{\rm WKB}$ is defined as
\begin{equation}
\label{WKBmode}
A_\lambda^{\rm WKB}(k,\tau)=\frac{1}{\sqrt{2 \Omega_{\lambda}(k \,, \tau)}}
e^{\imath\int^{\tau}\dd\tau'\, \Omega_{\lambda}(k, \tau')},
\end{equation}
with $\lambda=\pm$. The mass term regulator $\mu$ is inserted to avoid  additional IR divergences which are introduced by the adiabatic expansion for massless fields.  
Inserting the adiabatic solution \eqref{WKBmode} 
in Eq.~\eqref{eqn:adievoeq}, 
we then obtain the following exact equation for 
the WKB frequencies $\Omega_{\lambda}$
\begin{equation}
\label{WKB}
\Omega^{2}_{\lambda}(k, \tau) =
\bar{\Omega}_{\lambda}^2(k, \tau)
+ \frac{3}{4}\left(\frac{{\Omega}_{\lambda}'(k, \tau)}{{\Omega}_{\lambda}(k, \tau)}\right)^2 - \frac{{\Omega}_{\lambda}''(k,\tau)}{2{\Omega}_{\lambda}
(k,\tau)},
\end{equation}
with
\be 
\bar{\Omega}_{\lambda}^2(k, \tau)= \omega^2(k, \tau) - \lambda k g \phi'(\tau)
\ee
and
\begin{equation}
\omega^2 (k,\tau)=k^2+\mu^2a^2(\tau) \,.
\end{equation}

The usual procedure is then to solve Eq.~\eqref{WKB} iteratively, introducing an adiabatic parameter 
$\epsilon$ assigning a power of $\epsilon$ to each of the derivative with respect to $\tau$.
To arrive to order $2n$, we have then to do $n$ iterations. 
Finally, we have to further Taylor-expand $\bar{\Omega}_{\lambda}$ in power of $\epsilon$ around $\epsilon=0$ 
discarding all the resulting terms of adiabatic order larger than $2 n$ in the final result.

One can then mode expand $A_\lambda({\bf x},\tau )$ using the $n$-th adiabatic order  mode functions 
$A_\lambda^{(n)}(k,\tau)$ and the associated adiabatic creation and annihilation operators, and then define the 
$n$-th order adiabatic vacuum as $a_{\lambda,k}^{(n)} |0^{(n)}\rangle=0$ when $\tau\rightarrow-\infty$. 
In particular, in our case we have that $\omega'/\omega\rightarrow0$ for $\tau\rightarrow-\infty$. 
Thus, in this limit the adiabatic vacuum defined at any adiabatic order becomes essentially the adiabatic vacuum of infinite order which we call $|0\rangle_A$.

The adiabatic regularization is a procedure to remove the UV divergences and consists in subtracting from an expectation value its adiabatic counterpart. In practice, we will proceed by introducing an UV physical cutoff $\Lambda$ for the mode integral, performing the subtraction, and only after that we will send the cutoff to infinity. Namely, we have
\begin{equation}
\label{reg}
\langle {\cal O} \rangle_{\textup{reg}}=\lim_{\Lambda \rightarrow \infty} \left[\langle {\cal O} \rangle_{\Lambda}-\langle {\cal O} \rangle_{\textup{A},\Lambda}\right] \,, 
\end{equation}
where by ${\cal O}$ we denote a quadratic operator in the quantum fields, such as the energy density or the helicity. By $\langle {\cal O} \rangle_{\Lambda}$ we then mean the bare expectation value of these operators evaluated with a UV cutoff $\Lambda$, while
by $\langle {\cal O} \rangle_{\textup{A},\Lambda}$ we mean the expectation value of their adiabatic counterpart evaluated using the same UV cutoff $\Lambda$.

Considering the energy density and the helicity, the bare expectation values are those computed in the previous section for $\mu=0$ (see Eqs.~\eqref{eqn:integral2} and \eqref{eqn:integral}). 
Their adiabatic counterpart
is instead given by their corresponding integrals expressed in terms of the WKB mode functions of a given adiabatic order $n$. 
Namely, these are given by Eqs.~\eqref{eqn:integral2} and \eqref{eqn:integral} where we take the adiabatic solution in Eq.~\eqref{WKBmode} for the mode function using the solution of adiabatic order $n$  and expanding again up to order $n$. In the case under consideration, the fourth order adiabatic expansion is needed in order to remove the UV divergences from the bare integral.
We then obtain:
\begin{widetext}
\begin{align}
\frac{\langle {\bf E}^2 + {\bf B}^2\rangle_{\textup{A},\Lambda}}{2} =&
\int_{c}^{\Lambda a} \frac{dk\,k^2}{(2\pi)^2 a^4}\Bigg[
\label{eqn:intadienergy} 
\left(\frac{1}{2 \Omega_+}+\frac{1}{2 \Omega_-}\right) \omega^2(k,\tau) 
+\frac{\Omega_-}{2}
+\frac{\Omega_+}{2}+\frac{\epsilon^2 \Omega_-''}{8 \Omega_-^3}+\frac{\epsilon ^2 \Omega_+''}{8
	\Omega_+^3}\Bigg]\notag\\
=&\int_{c}^{\Lambda a} \frac{dk\,k^2}{(2\pi)^2 a^4}\Bigg[2 \omega+\frac{\epsilon^2 k^4 \xi ^2}{\tau ^2 \omega^5}+\frac{\epsilon^2 k^2 \mu^2 \xi ^2}{H^2 \tau ^4 \omega^5}+\frac{\epsilon^4 15 k^8 \xi ^4}{4 \tau ^4 \omega^{11}}
-\frac{\epsilon^4 3 k^8 \xi ^2}{4 \tau ^4 \omega^{11}} \notag\\
&+\frac{\epsilon^4 3 \mu^8}{64 H^8 \tau ^{12} \omega^{11}}+\frac{\epsilon^4 3 k^2 \mu^6}{4 H^6 \tau ^{10} \omega^{11}}
-\frac{\epsilon^4 15 k^4 \mu^4}{16 H^4 \tau ^8 \omega^{11}} 
+\frac{\epsilon^4 19 k^2 \mu^6 \xi ^2}{8 H^6 \tau ^{10} \omega^{11}}+\frac{\epsilon^4 \mu^4}{4 H^4 \tau ^6 \omega^5}\notag\\
&+\frac{\epsilon^4 67 k^4 \mu^4 \xi ^2}{8 H^4 \tau ^8
	\omega^{11}}+\frac{\epsilon^4 15 k^6 \mu^2 \xi ^4}{2 H^2 \tau ^6 \omega^{11}}
	+\frac{\epsilon^4 21 k^6 \mu^2 \xi ^2}{4 H^2 \tau ^6 \omega^{11}}+\frac{\epsilon^4 15 k^4 \mu^4 \xi ^4}{4 H^4 \tau ^8
	\omega^{11}}
\Bigg],
\end{align}	
\begin{align} 
\label{eqn:intadieb}
\langle{\bf E}\cdot{\bf B}\rangle_{A,\Lambda} = &-\int_{c}^{\Lambda a} \frac{dk\,k^3}{(2\pi)^2 a^4}\Bigg[\frac{\epsilon  \Omega_+'}{2 \Omega_+^2}-\frac{\epsilon  \Omega_-'}{2 \Omega_-^2}\Bigg]
=\int_{c}^{\Lambda a}  \frac{dk\,k^3}{(2\pi)^2a^4}\Biggl[-\frac{\epsilon^2 k^3 \xi }{\tau ^2 \omega^5}+\frac{\epsilon^2 2 k \mu ^2 \xi }{H^2 \tau ^4 \omega^5}+\frac{\epsilon^4 121 k^3 \mu ^4 \xi }{8 H^4 \tau ^8 \omega^{11}}\notag\\
&-\frac{\epsilon^4 15
	k^7 \xi ^3}{2 \tau ^4 \omega^{11}}+\frac{\epsilon^4 3 k^7 \xi }{2
	\tau ^4 \omega^{11}}+\frac{\epsilon^4 5 k^5 \mu ^2 \xi ^3}{2 H^2 \tau ^6 \omega^{11}}-\frac{\epsilon^423 k^5 \mu ^2 \xi }{H^2 \tau ^6 \omega^{11}}
	+\frac{\epsilon^4k \mu ^6 \xi }{4 H^6 \tau ^{10} \omega^{11}}+\frac{\epsilon^4 10 k^3 \mu ^4 \xi ^3}{H^4 \tau ^8 \omega^{11}}
\Biggr] \,.
\end{align}
\end{widetext}
In the above equations the IR $k$-cutoff $c$ is also considered as an alternative to the mass term regulator to cure the IR divergences which appear when considering the fourth order adiabatic expansion of a massless field.

As said, the fourth order adiabatic expansion of the mode functions is sufficient to generate the same UV 
divergences of the bare integrals in Eqs.~\eqref{eqn:Idiv} and \eqref{eqn:Jdiv}. However, the fourth order expansion also generates logarithmic IR divergences since gauge fields are massless. 
One way to avoid the IR divergence is to use the mass term in Eq.~\eqref{eqn:adievoeq} 
as an IR regulator\footnote{See Ref.~\cite{Ferreiro:2018oxx} for an interpretation of the mass regulator $\mu$ in terms of running the coupling constant.}. In this way we get for the energy density:
\begin{align}
\label{energyadifin}
\frac{\langle {\bf E}^2 + {\bf B}^2\rangle^\mu_{\textup{A},\Lambda}}{2}=&\frac{\Lambda^4}{8 \pi^2}+\frac{\Lambda^2 \xi^2 H^2}{8 \pi^2}\notag\\&+\frac{3 H^4 \xi ^2 \left(5 \xi ^2-1\right)\log(2 \Lambda/H) }{16 \pi^2 }\notag\\
&-\frac{H^4}{480\pi ^2}-\frac{H^4 \xi ^2 \left(23 \xi ^2-9\right)}{16 \pi ^2}\notag\\
&-\frac{3 H^4 \xi ^2 \left(5 \xi ^2-1\right) \log \left(\frac{\mu }{H}\right)}{16 \pi ^2} \,.
\end{align}

Analogously, for the helicity term we get:
\begin{align}
\label{helicityadifin}
-\langle{\bf E}\cdot{\bf B}\rangle^\mu_{A,\Lambda}=&\frac{ \Lambda ^2 \xi H^2}{8 \pi ^2} + \frac{3 H^4 \xi  \left(5 \xi ^2-1\right) \log \left(\frac{2 \Lambda }{H}\right)}{8 \pi ^2}\notag\\
&-\frac{3 H^4 \xi  \left(5 \xi ^2-1\right) \log \left(\frac{\mu }{H}\right)}{8 \pi ^2}\notag\\&+\frac{H^4 \left(19 \xi -56 \xi
	^3\right)}{16 \pi ^2}.
\end{align}

Note that our WKB ansatz correctly reproduce the the UV divergences of the energy density and helicity terms. As already known, the fourth order adiabatic expansions leads also to finite terms, including a term with a logarithmic dependence on the effective mass regulator; note that in Eqs.~\eqref{energyadifin} and \eqref{helicityadifin} all the terms in $\mu$ which are regular for $\mu \rightarrow 0$ are omitted.
Let us also comment on the term independent on $\xi$, i.e. $H^4/(480 \pi^2)$ in Eq.~\eqref{energyadifin}.
This term is generated by the fourth order adiabatic subtraction and  is connected to the conformal anomaly. 
The term we find corresponds to twice the result for a massless conformally coupled scalar field, i.e. 
$H^4/(480 \pi^2)=2\times H^4/(960 \pi^2)$ \cite{Bunch:1978gb,Birrell:1982ix}, as expected since the 
two physical states $A_\pm$ behave like two conformally coupled massless scalar fields for $\xi=0$ 
\footnote{It was recently shown in Ref.~\cite{Chu:2016kwv} that in case of photons the standard result 
$\langle T  \rangle /4 =-31 H^4/(480 \pi^2)$ can be obtained by adiabatic regularization only by including 
Faddeev-Popov fields. On the other hand, it is possible to get the same result without consideration of the Faddeev-Popov ghosts by, first, calculating the photon vacuum polarization in the closed static FLRW (Einstein) universe in which all geometric terms in the trace conformal anomaly become zero and a non-zero average photon energy density arises due to the Casimir effect~\cite{St76}, and then using the known form of the conserved vacuum polarization tensor in a conformally flat space-time that produces the $R_{\mu\nu}R^{\mu\nu}-\frac{1}{3}R^2$ term in the trace anomaly.  Note that the same procedure yields the correct answer for spins $s=0,\frac{1}{2}$, too.}. 

An alternative procedure to avoid IR divergences in the adiabatic subtraction is to introduce a time independent IR cutoff $k=c$ 
in the adiabatic integrals. 
In this way we obtain for the energy density and helicity term, respectively:
\begin{align}
\label{energyadifin_cutoff}
\frac{\langle {\bf E}^2 + {\bf B}^2\rangle^c_{\textup{A},\Lambda}}{2}=&\frac{\Lambda^4}{8 \pi^2}+\frac{\Lambda^2 \xi^2 H^2}{8 \pi^2}\notag\\&
+\frac{3 H^4 \xi ^2 \left(5 \xi ^2-1\right)\log(2 \Lambda/H) }{16 \pi^2 }\notag\\&-\frac{3 H^4 \xi ^2 \left(5 \xi ^2-1\right)\log\left(2 c/(a H)\right)}{16 \pi^2 }\notag\\&-\frac{c^4}{8 \pi^2 a^4}-\frac{c^2 \xi^2 H^2}{8 a^2 \pi^2}\,,
\end{align}

\begin{align}
\label{helicityadifin_cutoff}
-\langle{\bf E}\cdot{\bf B}\rangle^c_{A,\Lambda}=&\frac{ \Lambda ^2 \xi H^2}{8 \pi ^2} + \frac{3 H^4 \xi  \left(5 \xi ^2-1\right) \log \left(\frac{2 \Lambda }{H}\right)}{8 \pi ^2}\notag\\
&-\frac{ c^2 \xi H^2}{8 a^2 \pi^2} - \frac{3 H^4 \xi  \left(5 \xi ^2-1\right) \log \left(\frac{2 c}{a H}\right)}{8 \pi ^2}.
\end{align}
As for the case with the effective mass regulator, 
the UV divergences of the energy density and helicity terms are also correctly reproduced, although 
the finite terms are different. Let us note that the term connected to the conformal anomaly is 
absent, as it comes from the $k=0$ pole structure of the WKB energy density integrand 
\cite{Birrell:1982ix,Parker:2009uva}. By comparing Eqs.~\eqref{energyadifin_cutoff}-\eqref{helicityadifin_cutoff} with 
\eqref{energyadifin}-\eqref{helicityadifin}, the terms which do not depend on $\Lambda$ obtained with the IR cutoff can have the same time dependence of the effective mass term by instead considering a comoving IR cutoff $c=\Lambda_{\rm IR} a$. Further quantitive consistency from the two approaches can be obtained by matching 
$\mu$ to a physically motivated value for $\Lambda_{\rm IR}$ correspondent to the scale at which the WKB approximation breaks down (see \cite{Durrer:2009ii,Marozzi:2011da}).

We have seen that the adiabatic regularization applied to the two physical helicity states, because of 
the massless nature of the gauge field, introduces logarithmic IR divergences. This effect 
happens in many other contexts, see e.g. Ref.~\cite{Seery:2010kh} for a review
of IR effects in de Sitter space-time. 
Logarithmic IR divergences in averaged quantities also appear in the context
of the Schwinger effect in de Sitter where they lead to the so called
IR Hyper-conductivity effect \cite{Kobayashi:2014zza}. Furthermore,
it has been shown that such logarithmic IR divergences also
appear when using other renormalization methods such as point splitting
renormalization and thus seem to be generic and not specific  of the adiabatic regularization method  \cite{Hayashinaka:2016dnt}.

%%%%%%%%%%%%%%%%%%%%%%%%%%%%%%%%%%%%%%%%%%%%%%%%%%
\subsection{Counterterms}
\label{counterterms}
%%%%%%%%%%%%%%%%%%%%%%%%%%%%%%%%%%%%%%%%%%%%%%%%%%

We have introduced a fourth order adiabatic expansion which correctly reproduced the UV divergences of the bare quantities. These divergent terms are associated to non-renormalizable derivative interaction counter-terms of the pseudo-scalar field:
\begin{align}
\Delta\mathcal{L} = &- \frac{\alpha }{4}\nabla^\mu \nabla^\nu 
\phi\nabla_\mu \nabla_\nu 
\phi\notag\\
\label{eqn:Clagrangian}& - \frac{\beta }{4}\nabla^\mu \phi \nabla^\nu \phi \nabla_\mu \phi \nabla_\nu\phi,
\end{align}
where  $\alpha$ and $\beta$ are constants of mass dimension 
$-2$ and $-4$, respectively. 

With the new interaction added, the  Klein-Gordon equation for the inflaton becomes:
\be
\label{newkg}
[\square  - \alpha \square^2 + \beta (\nabla\phi)^2\square]\phi = V_{\phi}+\frac{g}{4}F^{\mu\nu}\tilde{F}_{\mu\nu},
\ee
where $\square^2\equiv\nabla^\mu \nabla^\nu\nabla_\mu \nabla_\nu$. 
The two additional terms in Eq.~\eqref{eqn:Clagrangian} lead to the following modification of the energy density:
\begin{align}
T_{00}^{(\phi)} =& \Lambda+\frac{{\phi'}^2}{2 a^2} + V(\phi) \notag\\
&+ \frac{\alpha}{a^4} \left(c_1{\phi''}^2+c_2\phi'\phi^{(3)}\right)  +\frac{\beta}{a^4} c_3{\phi'}^4 
\end{align}
where we have also added a cosmological constant $\Lambda$, and the values of the constants $c_i$, $i=1,2,3$, 
are not important for our purposes.

We now isolate the divergences coming from the energy density and the helicity integral using 
dimensional regularization \cite{Bunch:1980vc}, where, working in a generic $n$-dimensional FRLW 
space-time, the UV divergences show up as poles at $n=4$. This makes clear and explicit 
the connection between adiabatic expansion and counterterms. 
We will use in this Section results derived in the previous Section.
However, we 
will keep explicit track of derivatives of the pseudo-scalar field $\phi$ here, instead of using the 
variable $\xi$.

The integral measure in $n$ dimensions is:
\begin{equation}
\int \frac{\dd k\,k^2}{(2 \pi)^3}\to\int \frac{\dd k\,k^{n-2}}{(2\pi)^{n-1}}.
\end{equation}

%%%%%%%%%%%%%%%%%%%%%%%%%%%%%%%%%%%%%%
\subsubsection{Energy Density}
%%%%%%%%%%%%%%%%%%%%%%%%%%%%%%%%%%%%%%

As noted in Section~\ref{Energy-momentum tensor} and \ref{Analytical calculation}, the energy density 
of the gauge field presents quartic, quadratic and logarithmic divergences.
From the adiabatic expansion of Eq.~\eqref{eqn:integral2}, the term that contribute to the quartic 
divergence is:
\begin{align}
\int_{0}^{\infty}\frac{dk\,k^{n-2}}{(2\pi)^{n-1} a^4}2 \sqrt{k^2+ \mu^2a^2}=\notag\\
\label{key}
\frac{\mu^4}{16 \pi ^2 (n-4)}+\cdots,
\end{align}
where we have retained only the pole at $n=4$. Eq. \eqref{key} shows that the quartic divergence can 
be absorbed by the cosmological constant counterterm $\delta\Lambda$.
Similarly, the quadratic divergence comes from:
\begin{align}
\int_{0}^{\infty}\frac{dk\,k^{n-2}}{(2\pi)^{n-1} a^4}\frac{g^2 k^4 \phi '^2}{4 \left(k^2+\mu^2 a^2\right)^{5/2}}=\notag\\\frac{5 g^2 \mu^2 {\phi '}^2}{32 \pi ^2 a^2 (n-4)}+\cdots,
\end{align}  
which shows that we can absorb the quadratic divergence in the field strength counterterm $\delta Z$.
Finally, the logarithmic divergences come from the terms:
\begin{align}
&\int_{0}^{\infty}\frac{dk\,k^{n-2}}{(2\pi)^{n-1} a^4}\Bigg[\frac{15 g^4 k^8 \left(\phi '\right)^4}{64 \left(k^2+ \mu^2a^2\right)^{11/2}}\notag\\&+\frac{g^2 k^8 \left(\phi ''\right)^2}{16\left(k^2+ \mu^2a^2\right)^{11/2}}-\frac{g^2 k^8 \phi ^{(3)} \phi '}{8 \left(k^2+ \mu^2a^2\right)^{11/2}}\Bigg]=\notag\\&
-\frac{15 g^4 {\phi '}^4}{256 \pi ^2 (n-4) a^4}-\frac{g^2 (\phi '')^2}{64 \pi ^2 a^4 (n-4)}\notag\\&+\frac{g^2 \phi ^{(3)} \phi '}{32 \pi ^2 a^4 (n-4)}.
\end{align}
The first term can be absorbed in the counterterm $\delta\beta$, whereas the second and the third can 
be absorbed in the $\delta\alpha$ counterterm.

\subsubsection{Helicity Integral}

We now consider the divergences in the adiabatic approximation of $g\langle{\bf E} \cdot {\bf B}\rangle$, since this is the term which enters the Klein-Gordon equation~\eqref{newkg}, to see which are the counterterms needed to absorb them. The helicity integral contains 
only quadratic and logarithmic divergences.

The quadratic divergence comes from the term:
\begin{align}
\int_{0}^{\infty}\frac{\dd k\,k^{n-2}}{(2\pi)^{n-1} a^4}\frac{g^2 k^3 \phi ''}{2 \left(k^2+\mu^2a^2\right)^{5/2}}=\notag\\
\frac{5 g^2 \mu^2 \phi ''}{16 \pi ^2 a^2 (n-4)}+\cdots,
\end{align}
which, again, can be absorbed in the redefinition of the scalar field $\delta Z$. Note that the factor 
of $a^2$ at the denominator is not a problem since every term with the derivative of the scalar field 
in the Klein-Gordon equation~\eqref{newkg} contains it.

The logarithmic divergence comes instead from the terms:
\begin{align}
&\int_{0}^{\infty}\frac{dk\,k^{n-2}}{(2\pi)^{n-1} a^4}\Bigg[-\frac{15 g^4 k^7 {\phi '}^2 \phi ''}{16 \left(k^2+ \mu^2a^2\right)^{11/2}}\notag\\&+\frac{g^2 k^7 \phi ^{(4)}}{8 \left(k^2+ \mu^2 a^2\right)^{11/2}}\Bigg]=\notag\\
&\frac{g^4 {\phi '}^2 \phi ''}{4 \pi ^2 a^4 (n-4)}-\frac{g^2 \phi ^{(4)}}{32 \pi ^2 a^4 (n-4)}+\cdots,
\end{align}
which can be absorbed in the counterterms $\delta\beta$ and $\delta\alpha$ respectively.
%%%%%%%%%%%%%%%%%%%%%%%%%%%%%%%%%%%%%%%%%%%%%%%%%%%%%%%%%%%%%%

\section{Implications for background dynamics}
\label{implicationsdissipative}

\begin{figure}
        \centering
        \includegraphics[width=\columnwidth]{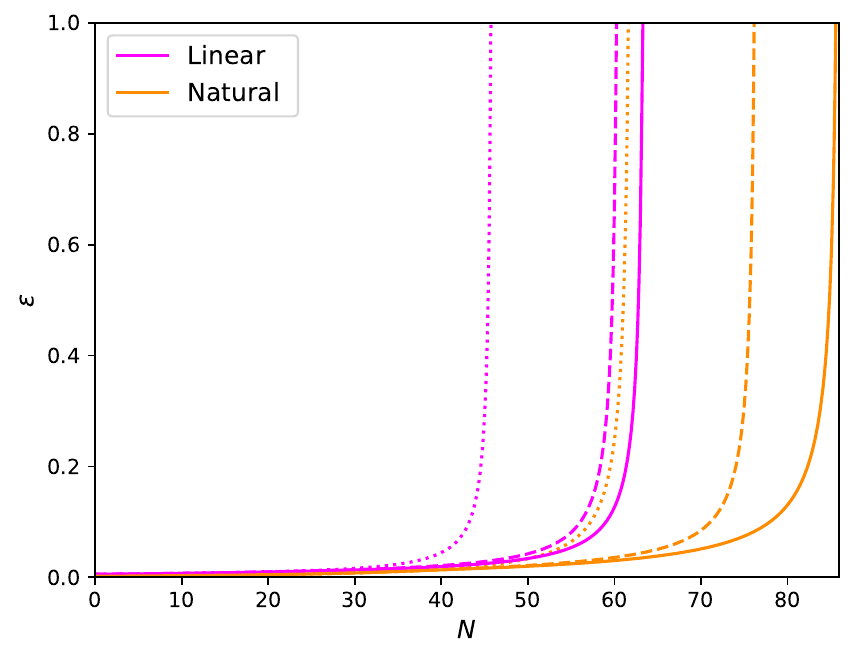} \\
        \includegraphics[width=\columnwidth]{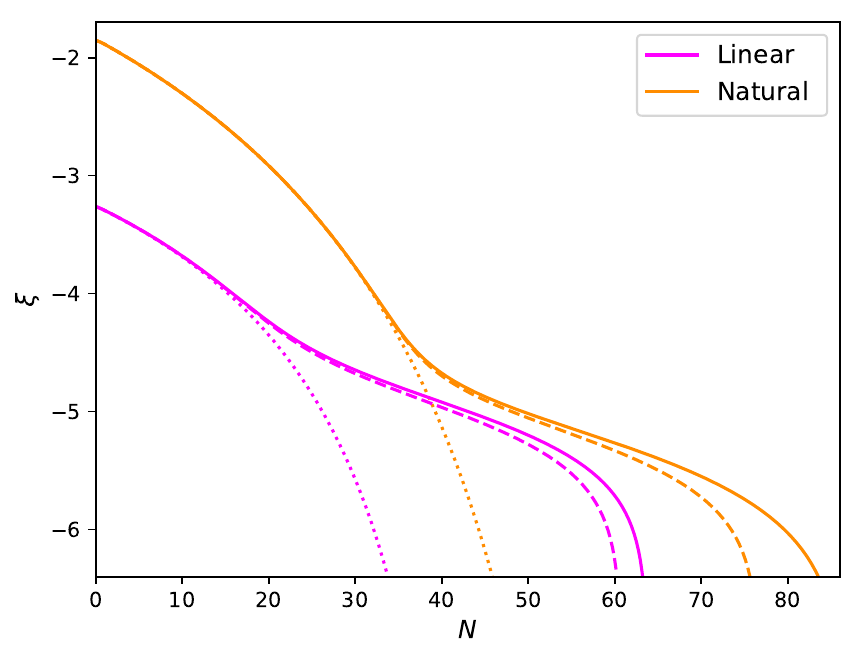}
        \caption{Numerical evolution of $- \dot H/H^2$ (top) and $\xi$ (bottom) for the 
        linear model (magenta line) and for natural inflation (orange line) for $|\xi| \sim 5$ (solid) 
        compared with the case by extrapolating $\xi \gg 1$ results (dashed) or no coupling (dotted). 
        We have chosen $\lvert g\rvert=60/M_{\textup{pl}}$ in both cases. For natural inflation we have 
        used $f=5 M_{\textup{pl}}$.}
        \label{fig:Numeric}
\end{figure}

\begin{figure}
	\centering
	\includegraphics[width=\columnwidth]{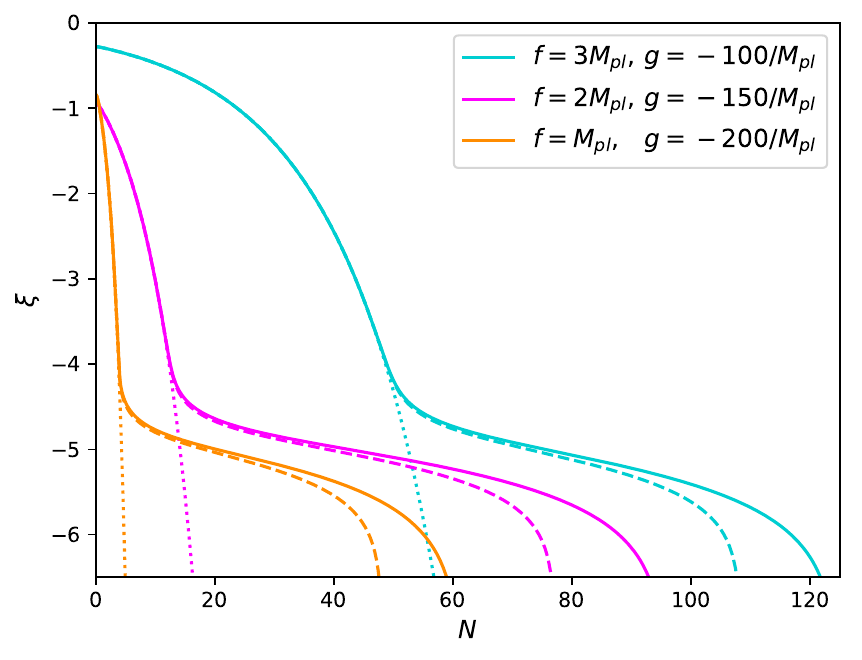} 
	\caption{Numerical evolution of  $\xi$ for  natural inflation for three different  values of $f$ and $g$. We use dotted lines for no backreaction, dashed lines for extrapolated results for backreaction and solid lines for our results.}
	\label{fig:NumericNatural}
\end{figure}

We now consider the implications of our results for the background dynamics.
The regularized helicity integral term behaves as an additional effective friction term and slows down 
the inflaton motion through energy dissipation into gauge fields. The regularized energy-momentum tensor 
of the gauge field produces an additional contribution to the Friedmann equations. 

In order to study backreaction, we introduce the quantity $\Delta$ which parameterizes the 
contribution of the gauge fields to the number of $e$-folds 
during inflation:
	\begin{align}
	N&=H\int\, \frac{d\phi}{\dot{\phi}} \simeq -\int\,d\phi\,
\frac{3 H^2}{V'}\left[1-g\frac{\langle {\bf E} \cdot {\bf B} \rangle}{3 H \dot{\phi}}\right]\notag\\
	\label{efolds}
	&\equiv\bar{N}(1+\Delta),
	\end{align}  
	where in the second equality we have used the Klein-Gordon equation during 
slow-roll and we have defined $\bar{N}$ as the number of $e$-folds without taking backreaction into account. For simplicity we consider the case of a minimal subtraction scheme to avoid the analysis for different values of the IR mass term regulator or cutoff involved in the adiabatic subtraction described in the previous section.

The extreme case with strong dissipation and strong coupling, i.e. $3 H \dot \phi \ll g \langle E \cdot B \rangle$ with $\xi \gg 1$,  
has been the target of the original study in \cite{Anber:2009ua}. For $\xi \gg 1$ our exact results for the averaged 
energy-momentum tensor and helicity term differ about 10 \% from 
the approximated ones in \cite{Anber:2009ua} and therefore we find estimates consistent with \cite{Anber:2009ua} 
at the same level of accuracy.

As we have shown, our results for $\xi \lesssim 10$ differ from previous ones in the literature. We can give an estimate of the difference in the number of $e$-folds. As a working example, we use a linear potential $V(\phi)=\Lambda^4(1-C\lvert\phi\rvert)$ and we compare our results obtained with those for $|\xi| \sim 5$ based on the incorrect extrapolation from $\xi \gg 1$ in Eqs.~\eqref{E2B2_literature} and 
\eqref{EB_literature}. Assuming a standard value of $H\sim 2\times 10^{-5} M_\textup{pl}$, a coupling $\sim 60$ and $\lvert\xi\rvert\sim 5$ we obtain $\Delta\simeq0.32$ and $0.37$ for the extrapolated and exact result respectively. Our results thus leads to a $5\%$ longer duration of inflation compared to the extrapolated ones in this case. We note that, when back-reaction changes the duration of inflation appreciably, it is possible that the gauge field contribution to $\dot H$ is not negligible when observationally relevant scales exit from the Hubble radius, potentially affecting the slopes of the primordial spectra.

To complement and confirm these 
analytic estimates we now present numerical results based
on the Einstein-Klein-Gordon equations \eqref{eqn:Friedmann}, \eqref{eqn:Einstein} and \eqref{KG} including the
averaged energy-momentum tensor and helicity of gauge fields where we allow $\xi=g \dot{\phi}/2 H$ to vary with time. In the case of the aforementioned linear potential, $V(\phi)=\Lambda^4(1-C |\phi|)$, and of $V(\phi) = \Lambda^4 \left[ 1 + \cos (\phi/f) \right]$ with $f \sim 2 M_\mathrm{pl}$ (such value of $f$ is close to the regime for which natural inflation 
is well approximated by a quadratic potential \cite{Savage:2006tr}).
The results are shown in Fig.~\ref{fig:Numeric}, comparing our exact results 
for $|\xi| \sim 5$ (solid) with those for $|\xi| \sim 5$ based on the incorrect 
extrapolation from $\xi \gg 1$ (dashed) and those in absence of gauge fields (dotted). As can be seen the approximation of $\xi\sim \textup{const}$ works very well in both the models.

\begin{figure}
	\centering
	\includegraphics[width=\columnwidth]{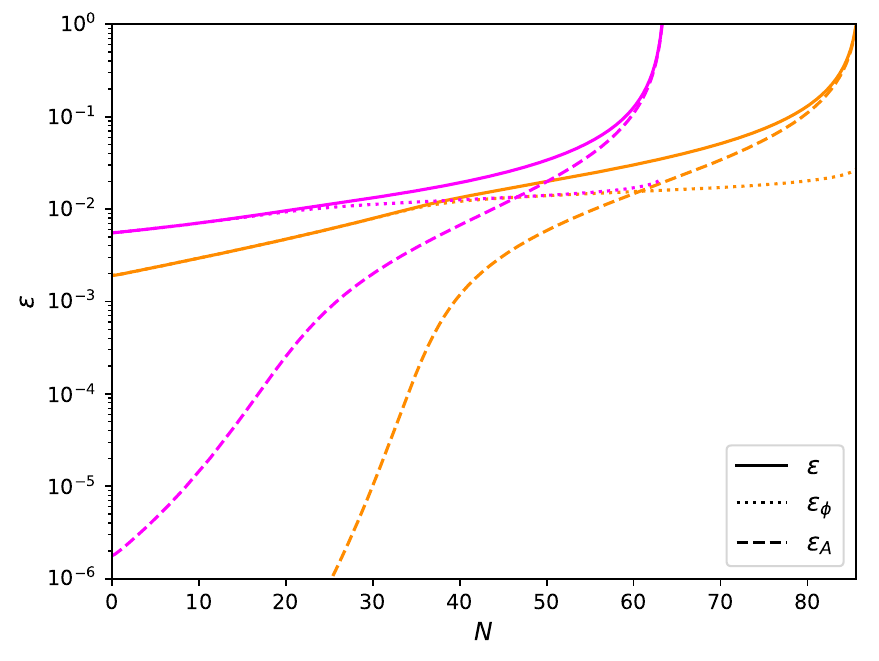} 
	\caption{Numerical evolution of  $\epsilon$, $\epsilon_\phi$ and $\epsilon_A$ as defined in the text for the same models shown in Fig.~\ref{fig:Numeric}.}
	\label{fig:NumericEpsilonDiff}
\end{figure}

Fig.~\ref{fig:NumericNatural} shows the importance of backreaction in the case of natural inflation for three different values of $f$.
The inflaton decay into gauge fields allows for a longer period of inflation compared to the case in which coupling to gauge fields is absent.   
Figs. \ref{fig:Numeric} and \ref{fig:NumericNatural} also show that our correct expressions lead to a longer period of inflation than the incorrect extrapolation from $\xi >>1$. Furthermore, we show in Fig.~\ref{fig:NumericEpsilonDiff} how the slow-roll parameter $\epsilon$ is dominated by $\epsilon_A\equiv \langle \mathbf{E}^2+\mathbf{B}^2 \rangle/3 M_\textup{pl}^2 H^2$ rather than by the usual scalar field contribution $\epsilon_\phi\equiv\dot{\phi}^2/2M_\textup{pl}^2$ at the end of inflation. 
Note also that the previously unexplored regime $\xi\ll 1$ is regular and included in our calculations
whereas the approximation of Eqs.~\eqref{E2B2_literature} and \eqref{EB_literature} become singular in this regime. 

%%%%%%%%%%%%%%%%%%%%%%%%%%%%%%%%%%%%%%%%%%%%%%%%%%%%%%
\section{Conclusions}
\label{Conclusions}
%%%%%%%%%%%%%%%%%%%%%%%%%%%%%%%%%%%%%%%%%%%%%%%%%%%%%%%%

We have studied the backreaction problem for a pseudo-scalar field $\phi$ which drives inflation and is coupled to gauge fields. As in other problems in QFT in curved space-times,
this backreaction problem is plagued by UV divergences.
We have identified the counterterms necessary to heal the UV divergences
for this not renormalizable interaction, which are higher order in scalar field derivatives.
We have also introduced a suitable adiabatic expansion capable to include the correct divergent terms of the integrated quantities.
Under the assumption of a constant time derivative of the inflaton, we have performed analytically the
Fourier integrals for the energy density and for the helicity in an exact way with an identification of divergent and finite terms.

Since previous approximate results were available only for $\xi \gg 1$,
our calculation which is valid for any $\xi$ has uncovered new aspects
of this backreaction problem.
We have shown that the regime of validity of previous results is $\xi
\gtrsim 10$ with a $10 \%$ level of accuracy.
We have then provided results which are more accurate than those
present in the literature in the regime $\xi \lesssim 10$.

Our results show that the inflaton decay into gauge fields leads to a
longer stage of inflation even for
$\xi \lesssim 10$. This is particularly relevant for natural inflation
since $f \lesssim M_{\textup{pl}}$ is a viable regime
for a controlled effective field theory \cite{ArkaniHamed:2003wu} and
a controlled limit of string theory \cite{Banks:2003sx}.

The techniques of integration used here in the computation of the bare
integrals of $\langle {\bf E}^2 \rangle$, $\langle {\bf B}^2 \rangle$ and $\langle {\bf
E}\cdot{\bf B} \rangle$
could have a wide range of applications in axion inflation, baryogenesis and magnetogenesis. Moreover, it would be interesting to
compute analytically the energy-momentum tensor and the helicity term
for a non-constant time derivative of the pseudo-scalar
field which we have adopted in this paper.
Other directions would be the calculation of the contribution
of gauge fields onto scalar fluctuations leading to non-Gaussian
corrections to the primordial power spectra and onto
gravitational waves at wavelengths which range from CMB observations
to those relevant for the direct detection from current and future
interferometers. We hope to return to these interesting topics in future works.

\begin{acknowledgments}
We wish to thank Marco Peloso for comments on the draft. We also thank Chiara Animali and Pietro Conzinu for independently checking several equations presented here. MBa, MBr, FF would like to thank INFN under the program InDark (Inflation, Dark Matter and Large Scale Structure) for financial support.
GM would like to thank INFN under the program TAsP (Theoretical Astroparticle Physics) for financial support. AAS was supported by the Russian Science Foundation grant 16-12-10401.
\end{acknowledgments}

%%%%%%%%%%%%%%%%%%%%%%%%%%%%%%%%%%%%%%%%%%%%%%%%%%%%

\setcounter{section}{0}

\begin{widetext}
	
	%%%%%%%%%%%%%%%%%%%%%%%%%%%%%%%%%%%%%%%%%%%%%%%%%%%%%%%%%%%%%%%%%%%%%%%%%%%%
	\subsection{Calculation of the energy density and its backreaction}
	\label{Appendix5}
	%%%%%%%%%%%%%%%%%%%%%%%%%%%%%%%%%%%%%%%%%%%%%%%%%%%%%%%%%%%%%%%%%%%%%%%%%%%%
	
	In this appendix we give the analytical expressions of the bare integrals in Eqs.~\eqref{eqn:integral2} and \eqref{eqn:integral}. We carefully show the calculation of the energy density of Eq.~\eqref{eqn:integral2}; the calculation  of Eq.~\eqref{eqn:integral} is then straightforward, so we only outline the differences from the one for the energy density. In the following we will use techniques introduced in Ref.~\cite{Kobayashi:2014zza,Frob:2014zka}.
	
	\vspace{1cm}

\begin{center}	
	\bf{Energy density}
\end{center}

	We write  Eq.~\eqref{eqn:integral2} as:
	\begin{equation}
	\frac{\langle {\bf E}^2 + {\bf B}^2 \rangle}{2} 
	\equiv \frac{1}{(2 \pi)^2 a^4}\lim_{\Lambda\to\infty}\left[\mathcal{I}(\xi,\tau,\Lambda)+\mathcal{I}(-\xi,\tau,\Lambda)\right]\, ,
	\end{equation}
	where 
	
	\begin{equation}
	\mathcal{I}(\pm\xi,\tau,\Lambda) 
	= \int_{0}^{\Lambda a}\,dk
	\,k^2\Bigl[k^2\left(|A_\pm|^2  \right) 
	+ |A_\pm'|^2  \Bigr]\, ,
	\end{equation}
	and  $\Lambda$ is an UV  physical cutoff (recall that the physical momentum $k_\textup{phys}$ is related to the comoving one by $k_\textup{com}=k_\textup{phys}a$) used to isolate the UV divergences.
	Using Eq.~\eqref{eqn:aplusaminus} and the properties of the Whittaker functions $(W_{\lambda\, ,\sigma}(x))^*=W_{\lambda^*\, ,\sigma^*}(x^*)$ and  $\frac{d}{dx}W_{\lambda\, ,\sigma}(x)=(\frac{1}{2}-\frac{\lambda}{x})W_{\lambda\, ,\sigma}(x)-\frac{1}{x}W_{\lambda+1\, ,\sigma}(x)$, we obtain:
	\begin{align}
	\label{Icsi+}
	\mathcal{I}(\xi,\tau,\Lambda)=\int_{0}^{\Lambda a}\,dk\,k^3 \frac{e^{\pi \xi}}{2}\Biggl\{\left[1+\left(1+\frac{\xi}{k \tau}\right)^2\right]W_{\imath \xi\,,\frac{1}{2}}(-2 \imath k \tau)    W_{-\imath \xi\,,\frac{1}{2}}(2 \imath k \tau) \notag\\+\left(\frac{\imath}{2 k \tau}+\frac{\imath\xi}{2 k^2 \tau^2}\right)\Bigl[  W_{\imath \xi,\frac{1}{2}}(-2 \imath k \tau)W_{-\imath \xi+1\,,\frac{1}{2}}(2 \imath k \tau)  
	-W_{-\imath \xi,\frac{1}{2}}(2 \imath k \tau)W_{\imath \xi+1\,,\frac{1}{2}}(-2 \imath k \tau) \Bigr]  \notag\\
	+\frac{1}{2 k^2 \tau^2}W_{-\imath \xi+1,\frac{1}{2}}(2 \imath k \tau)W_{\imath \xi+1\,,\frac{1}{2}}(-2 \imath k \tau)   \Biggr\}.
	\end{align}
	In order to solve this integral we now make use of the Mellin-Barnes representation of the Whittaker function $W_{\lambda,\sigma}(x)$:
	\begin{equation}
	\label{mellin}
	W_{\lambda,\sigma}(x)= 
	e^{-\frac{x}{2}}\int_{\mathcal{C}_s}\,\frac{ds}{2\pi\imath}\,\frac{\Gamma(-s+\sigma+\frac{1}{2})\Gamma(-s-\sigma+\frac{1}{2})\Gamma(s-\lambda)}{\Gamma(-\lambda-\sigma+\frac{1}{2})\Gamma(-\lambda+\sigma+\frac{1}{2})}x^s,
	\end{equation}
	with $\lvert \textup{arg}\,x\rvert<\frac{3}{2}\pi$ and the integration contour $\mathcal{C}_s$ runs from $-\imath \infty$ to $+\imath \infty$ and is chosen to separate the poles of $\Gamma(-s+\sigma+\frac{1}{2})$ and $\Gamma(-s-\sigma+\frac{1}{2})$ from those of $\Gamma(s-\lambda)$.
	
	Using Eqs. \eqref{Icsi+} and \eqref{mellin}, the reflection formula for the Gamma functions and integrating the $k$ dependent factor up to the cutoff $\Lambda$, it is straightforward to find
	\begin{equation}
	\label{eqn:Ipm}
	\mathcal{I}(\xi,\tau,\Lambda)+\mathcal{I}(-\xi,\tau,\Lambda)=\mathcal{I}_1+\mathcal{I}_2+\mathcal{I}_3\,,
	\end{equation}
	where
	\begin{align}
	\label{I1}
	\mathcal{I}_1&=\frac{\sinh^2(\pi\xi)}{2\pi^2}\int_{\mathcal{C}_s}\,\frac{ds}{2\pi\imath}\,\int_{\mathcal{C}_t}\,\frac{dt}{2\pi\imath}\,(2\imath\tau)^{s+t}\Gamma(-s)\Gamma(1-s)\Gamma(-t)\Gamma(1-t)\Biggl\{ \Biggl[\left(e^{\imath\pi(s-\imath\xi)}+e^{\imath\pi(t+\imath\xi)}\right) \frac{(a\Lambda)^{4+s+t}}{4+s+t} \notag\\
	&+  \left(e^{\imath\pi(s-\imath\xi)}-e^{\imath\pi(t+\imath\xi)}\right)\frac{\xi}{\tau} \frac{(a\Lambda)^{3+s+t}}{3+s+t} +  \left(e^{\imath\pi(s-\imath\xi)}+e^{\imath\pi(t+\imath\xi)}\right)\frac{\xi^2}{2\tau^2} \frac{(a\Lambda)^{2+s+t}}{2+s+t}  \Biggl]\Gamma(s-\imath\xi)\Gamma(t+\imath\xi)\,\notag\\
	&+\Biggl[\left(e^{\imath\pi(t-\imath\xi)}+e^{\imath\pi(s+\imath\xi)}\right) \frac{(a\Lambda)^{4+s+t}}{4+s+t} +  \left(e^{\imath\pi(t-\imath\xi)}-e^{\imath\pi(s+\imath\xi)}\right)\frac{\xi}{\tau} \frac{(a\Lambda)^{3+s+t}}{3+s+t} \notag\\
	&+  \left(e^{\imath\pi(t-\imath\xi)}+e^{\imath\pi(s+\imath\xi)}\right)\frac{\xi^2}{2\tau^2} \frac{(a\Lambda)^{2+s+t}}{2+s+t}  \Biggl]\Gamma(t-\imath\xi)\Gamma(s+\imath\xi)\Biggr\}\,,
	\end{align}

	\begin{align}
	\label{I2}
	\mathcal{I}_2&=\frac{\xi\sinh^2(\pi\xi)}{2\pi^2}\int_{\mathcal{C}_s}\,\frac{ds}{2\pi\imath}\,\int_{\mathcal{C}_t}\,\frac{dt}{2\pi\imath}\,(2\imath\tau)^{s+t}\Gamma(-s)\Gamma(1-s)\Gamma(-t)\Gamma(1-t) \notag\\
	&\times\Biggl\{ (1-\imath\xi)\Biggl[ \left(e^{\imath\pi(t-\imath\xi)}-e^{\imath\pi(s+\imath\xi)}\right)\frac{1}{\tau} \frac{(a\Lambda)^{3+s+t}}{3+s+t} +  \left(e^{\imath\pi(t-\imath\xi)}+e^{\imath\pi(s+\imath\xi)}\right)\frac{\xi}{\tau^2} \frac{(a\Lambda)^{2+s+t}}{2+s+t}  \Biggr]\Gamma(s+\imath\xi-1)\Gamma(t-\imath\xi) \notag\\
	&+(1+\imath\xi)\Biggl[ \left(e^{\imath\pi(s-\imath\xi)}-e^{\imath\pi(t+\imath\xi)}\right)\frac{1}{\tau} \frac{(a\Lambda)^{3+s+t}}{3+s+t} +  \left(e^{\imath\pi(s-\imath\xi)}+e^{\imath\pi(t+\imath\xi)}\right)\frac{\xi}{\tau^2} \frac{(a\Lambda)^{2+s+t}}{2+s+t}  \Biggr]\Gamma(s-\imath\xi-1)\Gamma(t+\imath\xi) \Biggr\} \,,
	\end{align}
	and
	
	\begin{align}
	\label{I3}
	\mathcal{I}_3&=\frac{(\xi^2+\xi^4)\sinh^2(\pi\xi)}{4\pi^2}\int_{\mathcal{C}_s}\,\frac{ds}{2\pi\imath}\,\int_{\mathcal{C}_t}\,\frac{dt}{2\pi\imath}\,\frac{(a\Lambda)^{2+s+t}}{2+s+t}\frac{1}{\tau^2}(2\imath\tau)^{s+t}\Gamma(-s)\Gamma(1-s)\Gamma(-t)\Gamma(1-t)\notag\\
	&\times\Biggl[ \left(e^{\imath\pi(t-\imath\xi)}+e^{\imath\pi(s+\imath\xi)}\right)\Gamma(s+\imath\xi-1)\Gamma(t-\imath\xi-1)+\left(e^{\imath\pi(s-\imath\xi)}+e^{\imath\pi(t+\imath\xi)}\right)\Gamma(t+\imath\xi-1)\Gamma(s-\imath\xi-1)\Biggr]\,,
	\end{align}
	where we have assumed $\Re(n+s+t)>0$ for the terms proportional to $\Lambda^{n+s+t}$ in order to have convergence.

	We now analyze each of these contributions in turn, starting from the first integral in Eq. \eqref{I1}. We integrate first over the variable $t$. Let us further specify the integration contour by requiring $\Re(t),\,\Re(s)<0$. The integrand can have poles at $t=\pm\imath\xi-n$ ($n=0,1,2,\dots$), located on the left of the integration contour of $t$, and at $t=n$ and $t=-4-s,\,-3-s,\,-2-s$, located on the right of the integration contour of $t$. We close the contour counterclockwise, on the left half-plane. The added contours do not contribute to the result since an integral of the integrand over a finite path along the real direction vanishes at $\Im(t)\to\pm \infty$ and because any integral in the region $\Re(t)<-5$ vanishes in the limit $\Lambda\to\infty$. The integral is thus $2\pi\imath$ times the sum of the residues of the poles:
	\begin{equation}
	t=\pm \imath\xi,\,\pm \imath\xi-1,\,\pm \imath\xi-2,\,\pm \imath\xi-3,\,\pm \imath\xi-4,\,\pm \imath\xi-5,\,-s-4,-s-3,-s-2.
	\end{equation}
	
	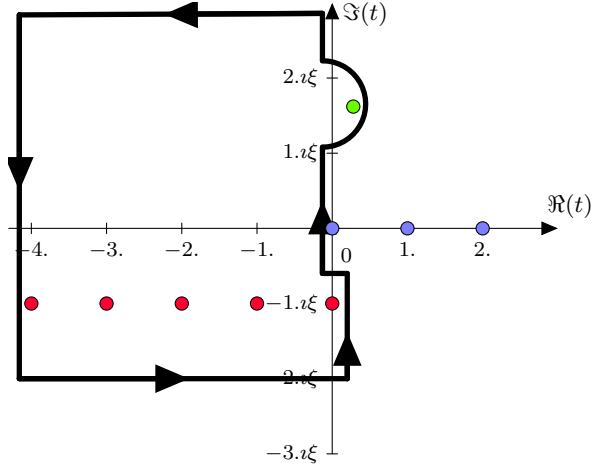
\begin{figure}[!ht]
    \centering
    \begin{minipage}{0.5\textwidth}
		\begin{tikzpicture}[line cap=round,line join=round,>=triangle 45,x=1.0cm,y=1.0cm]
		\draw[->,color=black] (-4.3,0.) -- (3.,0.);
		\foreach \x in {-4.,-3.,-2.,-1.,1.,2.}
		\draw[shift={(\x,0)},color=black] (0pt,2pt) -- (0pt,-2pt) node[below] {\footnotesize $\x$};
		\draw[color=black] (2.746362649294245,0.02702702702702703) node [anchor=south west] { $\Re{(t)}$};
		\draw[->,color=black] (0.,-3.) -- (0.,3.);
		\foreach \y in {-3.,-2.,-1.,1.,2.}
		\draw[shift={(0,\y)},color=black] (2pt,0pt) -- (-2pt,0pt) node[left] {\footnotesize $\y\imath\xi$};
		\draw[color=black] (0.019815418023887078,2.8513513513513513) node [anchor=west] { $\Im{(t)}$};
		\draw[color=black] (0pt,-10pt) node[right] {\footnotesize $0$};
		\clip(-4.3,-3.) rectangle (3.,3.);
		\draw [line width=2.pt] (-0.13,-0.6)-- (0.2,-0.6);
		\draw [line width=2.pt] (0.2,-0.6)-- (0.2,-2.);
		\draw [line width=2.pt] (0.2,-2.)-- (-4.16,-2.);
		\draw [line width=2.pt] (-4.16,-2.)-- (-4.16,2.84);
		\draw [line width=2.pt] (-4.16,2.84)-- (-0.13,2.86);
		\draw [line width=2.pt] (-0.13,2.86)-- (-0.13,2.23);
		\draw [line width=2.pt] (-0.13,-0.6)-- (-0.13,1.08);
		\draw [shift={(-0.13,1.655)},line width=2.pt]  plot[domain=-1.5707963267948966:1.5707963267948966,variable=\t]({1.*0.575*cos(\t r)+0.*0.575*sin(\t r)},{0.*0.575*cos(\t r)+1.*0.575*sin(\t r)});
		\draw [->,line width=2.pt] (-0.13,2.86) -- (-2.2341861198738173,2.849557388983256);
		\draw [->,line width=2.pt] (-4.16,2.84) -- (-4.16,0.5);
		\draw [->,line width=2.pt] (-4.16,-2.) -- (-1.9028318584070796,-2.);
		\draw [->,line width=2.pt] (0.2,-2.) -- (0.2,-1.38);
		\draw [->,line width=2.pt] (-0.13,-0.6) -- (-0.13,0.3684705882352941);
		\begin{scriptsize}
		\draw [fill=xdxdff] (0.,0.) circle (2.5pt);
		\draw [fill=xdxdff] (1.,0.) circle (2.5pt);
		\draw [fill=xdxdff] (2.,0.) circle (2.5pt);
		\draw [fill=ffqqtt] (-4.,-1.) circle (2.5pt);
		\draw [fill=ffqqtt] (-3.,-1.) circle (2.5pt);
		\draw [fill=ffqqtt] (-2.,-1.) circle (2.5pt);
		\draw [fill=ffqqtt] (-1.,-1.) circle (2.5pt);
		\draw [fill=ffqqtt] (0.,-1.) circle (2.5pt);
		\draw [fill=wwffqq] (0.28,1.62) circle (2.5pt);
		\end{scriptsize}
		\end{tikzpicture}
		\end{minipage}
		\caption{Integration contour $\mathcal{C}_t$ for the term proportional to $\Lambda^{4+s+t}\Gamma(t+\imath\xi)$ of Eq.~\eqref{I1}.  Blue points are poles of $\Gamma(1-t)\Gamma(-t)$ and lye outside the integration contour. Red points are poles of $\Gamma(t+\imath\xi)$ and lie inside the contour (in the terms proportional to $\Gamma(t-\imath\xi)$ the red points are in $t=\imath\xi-n$). The green point is the pole $t=-s-4$ and it  has been drawn there to emphasize that it slightly on the right of the imaginary axis (in the terms proportional to $\Lambda^{3+s+t}$ and $\Lambda^{2+s+t}$ the green point corresponds to $t=-s-3,\,-s-2$). The contour does not pass through any of the poles.}
		\label{contour1}
	\end{figure}
	
	Note that the poles at $t=-s-4,\,-s-3,\,-s-2$ lye slightly on the right of the axis $\Re(t)=0$, thus we slightly deform the contour to pick it up; the integration contour is illustrated in Fig.~\ref{contour1}. The latter poles give a contribution which is independent on the cutoff, whereas the sum of the former ones give a cutoff dependent results, we summarize this writing $\mathcal{I}_1$ as
	\begin{equation}
	\mathcal{I}_1=\mathcal{I}_{1,\Lambda}+\mathcal{I}_{1,\text{fin}}.
	\end{equation}
	
	We first analyze the cutoff dependent part of the result $\mathcal{I}_{1,\Lambda}$, that we summarize as follows in order to reduce clutter:
	\begin{equation}
	\mathcal{I}_{1,\Lambda}=\int_{\mathcal{C}_s}\,\frac{ds}{2\pi\imath}\Gamma(1-s)\Gamma(-s)\Biggl\{\Gamma(s-\imath\xi)\left[\mathcal{O}_1(\Lambda^{4+s-\imath\xi},\dots,\Lambda^{-1+s-\imath\xi})\right]
	+\Gamma(s+\imath\xi)\left[\mathcal{O}_2(\Lambda^{4+s+\imath\xi},\dots,\Lambda^{-1+s+\imath\xi})\right]\Biggr\}\,.
	\end{equation}
	The integral over $s$ can be carried out in the same way as the $t$ integral and is thus $2\pi \imath$ times the sum over the residues of the integrand from the points:
	\begin{equation}
	s= \pm \imath\xi,\,\pm \imath\xi-1,\,\pm \imath\xi-2,\,\pm \imath\xi-3,\,\pm \imath\xi-4\, , 
	\end{equation} 
	the residues from the points $s=\pm\imath\xi-n$ with $n>4$ vanish as $\Lambda\to\infty$. We schematically write the result of this integral as:
	\begin{equation}
	\mathcal{I}_{1,\Lambda}=f_{4}(\xi,\tau)\Lambda^4+f_{2}(\xi,\tau)\Lambda^2+f_{\textup{log}}(\xi,\tau)\log(2\Lambda/H)+f_1(\xi,\tau)
	\end{equation}
	and we will write  explicitly only the final result, together with the results from the integral $\mathcal{I}_2$ and $\mathcal{I}_3$.

	We now turn to calculate $\mathcal{I}_{1,\text{fin}}$, which is the sum of the pole of the integrand in the points $t=-4-s,\,-3-s,\,-2-s$ and can be written as:
	\begin{equation}
	\label{I1fin}
	\mathcal{I}_{1,\text{fin}}=\mathcal{I}_{1,t=-4-s}+\mathcal{I}_{1,t=-3-s}+\mathcal{I}_{1,t=-2-s}.
	\end{equation}
	We analyze in detail the integral over the  pole $t=-4-s$, the other two are similiar. The former is given by:
	\begin{equation}
	\label{finita}
	\mathcal{I}_{1,t=-4-s}=\int_{\mathcal{C}_s}\,\frac{ds}{2 \pi \imath}\,\frac{\pi \sinh ^2(\pi  \xi) }{64 \tau ^4}  \Biggl\{ \frac{\left(e^{-\pi  \xi -i \pi  s}+e^{ \pi  \xi + i \pi  s}\right)}{ \sin ^2(\pi  s) \sin (\pi  (s-i \xi ))}\Biggl[\frac{a_r}{s-\imath \xi}+B_r(s)-B_r(s-1)\Biggr]\Biggr\}+\xi\to-\xi\,,
	\end{equation}
	where $\xi\to-\xi$ stands for a second integral equal to the first one, but with $\xi$ replaced by $-\xi$ and $B_r(s)$ is a function of the form
	\begin{equation}
	B_r(s)=\frac{b_{r,1}}{s-\imath\xi+1}+\frac{b_{r,2}}{s-\imath\xi+2}+\frac{b_{r,3}}{s-\imath\xi+3}+\frac{b_{r,4}}{s-\imath\xi+4}+b_{r,5}s+b_{r,6}s^2+b_{r,7}s^3+b_{r,8}s^4
	\end{equation}
	and the coefficients $a_r,\,b_{r,j}$ for $j=1,\dots,8$ are independent on $s$.
	
	We first consider the term with $a_r$. We rewrite the integrand as follows:
	\begin{equation}
	\lim_{p\to 1}\frac{\pi \sinh ^2(\pi  \xi) }{64 \tau ^4} \frac{\left(e^{-\pi  \xi -i \pi  s}+e^{ \pi  \xi + i \pi  s}\right)}{ \sin ^2(\pi  s) \sin (\pi  (s-i \xi ))}\frac{a_r}{(s-\imath \xi)^p},
	\end{equation}
	with $p>1$. The integral of this function vanishes on an arc of infinite radius on the left half-plane, so we can close the contour on the left half-plane with a counterclockwise semicircle of infinite radius, as illustrated in Fig. \ref{contour2}. The integral is then $2\pi i$ times the sum of the residues in the poles $s=\pm\imath\xi-n$ and $s=-n-1$ with $n=0,\,1,\,2,\,\dots$. As for the other integrals we do not give the result here, but we will just write the final result.
	
	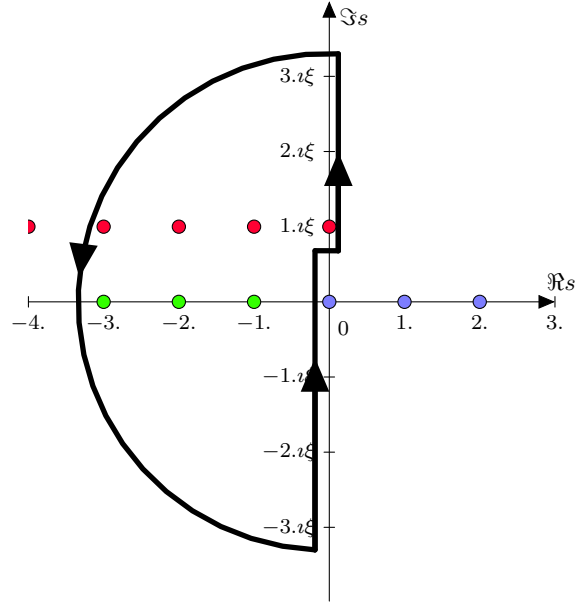
\begin{figure}[!ht]
	\centering
    \begin{minipage}{0.5\textwidth}
		\begin{tikzpicture}[line cap=round,line join=round,>=triangle 45,x=1.0cm,y=1.0cm]
		\draw[->,color=black] (-4.,0.) -- (3.,0.);
		\foreach \x in {-4.,-3.,-2.,-1.,1.,2.,3.}
		\draw[shift={(\x,0)},color=black] (0pt,2pt) -- (0pt,-2pt) node[below] {\footnotesize $\x$};
		\draw[color=black] (2.776960784313726,0.036951501154734404) node [anchor=south west] { $\Re{s}$};
		\draw[->,color=black] (0.,-3.9815242494226313) -- (0.,4.);
		\foreach \y in {-3.,-2.,-1.,1.,2.,3.}
		\draw[shift={(0,\y)},color=black] (2pt,0pt) -- (-2pt,0pt) node[left] {\footnotesize $\y\imath \xi$};
		\draw[color=black] (0.02144607843137255,3.7967667436489605) node [anchor=west] { $\Im{s}$};
		\draw[color=black] (0pt,-10pt) node[right] {\footnotesize $0$};
		\clip(-4.,-3.9815242494226313) rectangle (3.,4.);
		\draw [line width=2.pt] (0.12,3.3)-- (0.12,0.67949);
		\draw [line width=2.pt] (0.12,0.67949)-- (-0.19,0.68);
		\draw [line width=2.pt] (-0.19,0.68)-- (-0.19,-3.3);
		\draw [shift={(-0.035,0.)},line width=2.pt]  plot[domain=1.5238611249462244:4.665453778536017,variable=\t]({1.*3.303638146044448*cos(\t r)+0.*3.303638146044448*sin(\t r)},{0.*3.303638146044448*cos(\t r)+1.*3.303638146044448*sin(\t r)});
		\draw [->,line width=2.pt] (-0.19,-3.3) -- (-0.19,-0.7476238200675911);
		\draw [->,line width=2.pt] (0.12,0.67949) -- (0.12,1.989745);
		\draw [->,line width=2.pt] (-3.3022831900001357,0.4887592007753261) -- (-3.322684268078719,0.32427943725696745);
		\begin{scriptsize}
		\draw [fill=xdxdff] (0.,0.) circle (2.5pt);
		\draw [fill=xdxdff] (1.,0.) circle (2.5pt);
		\draw [fill=xdxdff] (2.,0.) circle (2.5pt);
		\draw [fill=ududff] (5.,0.) circle (2.5pt);
		\draw [fill=ttffqq] (-7.,0.) circle (2.5pt);
		\draw [fill=ttffqq] (-6.,0.) circle (2.5pt);
		\draw [fill=ttffqq] (-5.,0.) circle (2.5pt);
		\draw [fill=ttffqq] (-3.,0.) circle (2.5pt);
		\draw [fill=ttffqq] (-2.,0.) circle (2.5pt);
		\draw [fill=ttffqq] (-1.,0.) circle (2.5pt);
		\draw [fill=ffqqtt] (-1.,1.) circle (2.5pt);
		\draw [fill=ffqqtt] (-2.,1.) circle (2.5pt);
		\draw [fill=ffqqtt] (0.,1.) circle (2.5pt);
		\draw [fill=ffqqtt] (-3.,1.) circle (2.5pt);
		\draw [fill=ffqqtt] (-4.,1.) circle (2.5pt);
		\draw [fill=ffqqtt] (-5.,-1.) circle (2.5pt);
		\draw [fill=ffqqtt] (-6.,-1.) circle (2.5pt);
		\draw [fill=ffqqtt] (-7.,-1.) circle (2.5pt);
		\end{scriptsize}
		\end{tikzpicture}
		\end{minipage}
		\caption{Integration contour for the term proportional to $a_r$ in Eq.~\eqref{finita}. The radius of the semicircle is taken to be infinite and the contour does not pass through any of the poles. Blue points are the poles of $\Gamma(1-s)\Gamma(-s)$ whereas red points are the poles of $\Gamma(s-\imath\xi)$ and $\csc((s-\imath\xi))$. Green points are the poles of $\csc(\pi s)$. For the term with $\xi\to-\xi$ in Eq.~\eqref{finita} the red points move to $s=-\imath\xi-n$. }
		\label{contour2}
	\end{figure}
	Now we conclude integrating the terms with $B_r(s)-B_r(s-1)$. We shift the integration variable in the second term by $s\to y=s-1$ so that  the integral is given by:
	\begin{equation}
	\label{sandwich}
	\int_{\mathcal{C}_s}\,\frac{ds}{2 \pi \imath}\,(\dots) \left[B_r(s)-B_r(s-1)\right]=\left(\int_{\mathcal{C}_s}-\int_{\mathcal{C}_s-1}\right)\,\frac{ds}{2 \pi \imath}\,(\dots )B_r(s)\,,
	\end{equation}
	as illustrated in Fig.~\ref{contour2}.
	Thus we can evaluate this integral summing $2\pi\imath$ times the residues of the  singularities of the integrand which fall in the region sandwiched by the original integration contour and the shifted one, which are the poles at $s=-1$ and $s=\pm\imath\xi$. We  write the result as $F_1(\xi,\tau)$.
	
	\begin{figure}[!ht]
	\centering
    \begin{minipage}{0.5\textwidth}
		\begin{tikzpicture}[line cap=round,line join=round,>=triangle 45,x=1.0cm,y=1.0cm]
		\draw[->,color=black] (-4.,0.) -- (3.,0.);
		\foreach \x in {-4.,-3.,-2.,-1.,1.,2.,3.}
		\draw[shift={(\x,0)},color=black] (0pt,2pt) -- (0pt,-2pt) node[below] {\footnotesize $\x$};
		\draw[color=black] (2.7254901960784315,0.036951501154734404) node [anchor=south west] { $\Re{(s)}$};
		\draw[->,color=black] (0.,-3.9815242494226313) -- (0.,4.);
		\foreach \y in {-3.,-2.,-1.,1.,2.,3.}
		\draw[shift={(0,\y)},color=black] (2pt,0pt) -- (-2pt,0pt) node[left] {\footnotesize $\y\imath \xi$};
		\draw[color=black] (0.02144607843137255,3.7967667436489605) node [anchor=west] { $\Im{(s)}$};
		\draw[color=black] (0pt,-10pt) node[right] {\footnotesize $0$};
		\clip(-4.,-3.9815242494226313) rectangle (3.,4.);
		\draw [line width=2.pt] (0.15,3.3)-- (0.15,0.68);
		\draw [line width=2.pt] (0.15,0.68)-- (-0.1,0.68);
		\draw [line width=2.pt] (-0.1,0.68)-- (-0.1,-3.3);
		\draw [->,line width=2.pt] (0.15,0.68) -- (0.15,1.99);
		\draw [line width=2.pt] (0.15,3.298631726763146)-- (-0.95,3.3);
		\draw [line width=2.pt] (-0.95,3.3)-- (-0.95,0.68);
		\draw [line width=2.pt] (-0.95,0.68)-- (-1.1,0.68);
		\draw [line width=2.pt] (-1.1,0.68)-- (-1.1,-3.3);
		\draw [line width=2.pt] (-1.1,-3.3)-- (-0.1,-3.3);
		\draw [->,line width=2.pt] (0.15,3.298631726763146) -- (-0.3937966702683996,3.2993081471542536);
		\draw [->,line width=2.pt] (-0.95,3.3) -- (-0.95,1.7859296482412061);
		\draw [->,line width=2.pt] (-1.1,0.68) -- (-1.1,-1.5988364508224175);
		\draw [->,line width=2.pt] (-1.1,-3.3) -- (-0.6051247771836007,-3.3);
		\draw [->,line width=2.pt] (-0.1,-3.3) -- (-0.1,-1.1778432029335546);
		\begin{scriptsize}
		\draw [fill=xdxdff] (0.,0.) circle (2.5pt);
		\draw [fill=xdxdff] (1.,0.) circle (2.5pt);
		\draw [fill=xdxdff] (2.,0.) circle (2.5pt);
		\draw [fill=ududff] (5.,0.) circle (2.5pt);
		\draw [fill=ttffqq] (-7.,0.) circle (2.5pt);
		\draw [fill=ttffqq] (-6.,0.) circle (2.5pt);
		\draw [fill=ttffqq] (-5.,0.) circle (2.5pt);
		\draw [fill=ttffqq] (-3.,0.) circle (2.5pt);
		\draw [fill=ttffqq] (-2.,0.) circle (2.5pt);
		\draw [fill=ttffqq] (-1.,0.) circle (2.5pt);
		\draw [fill=ffqqtt] (-1.,1.) circle (2.5pt);
		\draw [fill=ffqqtt] (-2.,1.) circle (2.5pt);
		\draw [fill=ffqqtt] (0.,1.) circle (2.5pt);
		\draw [fill=ffqqtt] (-3.,1.) circle (2.5pt);
		\draw [fill=ffqqtt] (-4.,1.) circle (2.5pt);
		\draw [fill=ffqqtt] (-5.,-1.) circle (2.5pt);
		\draw [fill=ffqqtt] (-6.,-1.) circle (2.5pt);
		\draw [fill=ffqqtt] (-7.,-1.) circle (2.5pt);
		\end{scriptsize}
		\end{tikzpicture}
		\end{minipage}
		\caption{Integration contour in Eq.~\eqref{sandwich}.The contour does not pass through any of the poles. Blue points are the poles of $\Gamma(1-s)\Gamma(-s)$ whereas red points are the poles of $\Gamma(s-\imath\xi)$ and $\csc((s-\imath\xi))$. Green points are the poles of $\csc(\pi s)$. For the term with $\xi\to-\xi$ in Eq.~\eqref{finita} the red points move to $s=-\imath\xi-n$. }
		\label{contour3}
	\end{figure}
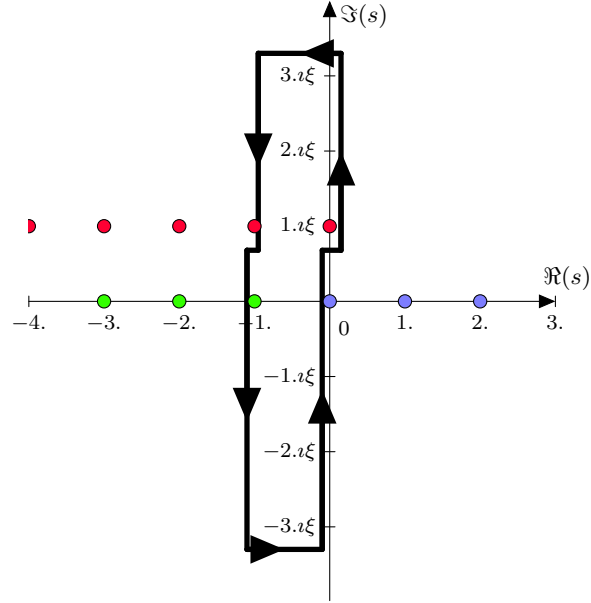
	The integral over the poles $t=-s-3,\,-s-2$ can be written in a similar way as:
	\begin{equation}
	\label{i13}
	\mathcal{I}_{1,t=-3-s}=\int_{\mathcal{C}_s}\,\frac{ds}{2 \pi \imath}\,\frac{\mathcal{A}_3(\xi,\tau)}{\sin^2(\pi s)\sin(\pi(s-\imath\xi))}\Biggl[\frac{c_r}{s-\imath \xi}+D_r(s)-D_r(s-1)\Biggr]+\xi\to-\xi\,,
	\end{equation}
	for the integral over the pole $t=-3-s$ and
	\begin{equation}
	\label{i12}
	\mathcal{I}_{1,t=-2-s}=\int_{\mathcal{C}_s}\,\frac{ds}{2 \pi \imath}\,\frac{\mathcal{A}_2(\xi,\tau)}{\sin^2(\pi s)\sin(\pi(s-\imath\xi))}\Biggl[\frac{e_r}{s-\imath \xi}+K_r(s)-K_r(s-1)\Biggr]+\xi\to-\xi\,,
	\end{equation}
	for the integral over the pole $t=-2-s$. The functions $\mathcal{A}_2$ and $\mathcal{A}_3$ are regular functions of $\xi$. $D_r(s)$ is given by:
	\begin{equation}
	D_r(s)=\frac{d_{r,1}}{s-\imath\xi+1}+\frac{d_{r,2}}{s-\imath\xi+2}+\frac{d_{r,3}}{s-\imath\xi+3}
	+d_{r,4}s+d_{r,5}s^2+d_{r,6}s^3
	\end{equation}
	and $K_r(s)$ is given by
	\begin{equation}
	K_r(s)=\frac{k_{r,1}}{s-\imath\xi+1}+\frac{k_{r,2}}{s-\imath\xi+2}
	+k_{r,3}s+k_{r,4}s^2
	\end{equation}
	The integrals in Eqs. \eqref{i13} and \eqref{i12} can be made exactly as done in the previous case for the integral over the residue in $t=-s-4$ in Eq.~\eqref{finita}.
	The full solution of the integral in Eq.~\eqref{I1} is thus:
	\begin{equation}
	\mathcal{I}_1=f_{4}(\xi,\tau)\Lambda^4+f_{2}(\xi,\tau)\Lambda^2+f_{\textup{log}}(\xi,\tau)\log(2\Lambda/H)+f_1(\xi,\tau)+f_{\textup{fin}}(\xi,\tau)\,,
	\end{equation}
	where we have included the contributions of Eqs.~\eqref{finita}, \eqref{i13} and \eqref{i12} in the term $f_{\textup{fin}}(\xi,\tau)$.
	
	The integrals $\mathcal{I}_2$ and $\mathcal{I}_3$ can be done following the same  procedure and we only give the final result in the following \footnote{Note that $\mathcal{I}_2$ is not explicitly invariant under the exchange of $t$ with $s$: we therefore symmetrize it before taking the integral in $s$ and $t$.}.

		Defining the contributions to the divergences and finite part of Eq.~\eqref{eqn:Ipm} as
		\begin{equation}
	\mathcal{I}(\xi,\tau,\Lambda)+\mathcal{I}(-\xi,\tau,\Lambda)=g_{4}(\xi,\tau)\Lambda^4+g_{2}(\xi,\tau)\Lambda^2+g_{\textup{log}}(\xi,\tau)\log(2\Lambda/H)+g_{\textup{fin}}(\xi,\tau),
	\end{equation}
    it can be found that the coefficient of the quartic divergence is:
	\begin{equation}
	g_4(\xi,\tau)=\frac{1}{2 H^4 \tau^4}\,.
	\end{equation}
	The coefficient of the quadratic divergence is:
	\begin{equation}
	g_2(\xi,\tau)=\frac{\xi^2}{2 H^2 \tau^4}\,.
	\end{equation}
	The coefficient of the logarithmic divergence is:
	\begin{equation}
	g_{\text{log}}(\xi,\tau)=\frac{3 \xi ^2 \left(5 \xi ^2-1\right) }{4 \tau ^4}\,.
	\end{equation}
	The finite part is
	\begin{align}
	g_{\textup{fin}}(\xi,\tau)=&\frac{-3 \xi ^2 \left(5 \xi ^2-1\right) (\psi (-\imath \xi -1)+\psi (\imath \xi -1))}{8 \tau ^4}+\frac{\xi ^2 \left(-79 \xi^4+22\xi^2+29\right)}{16 \left(\xi ^2+1\right) \tau ^4}\notag\\
	&+\frac{3 \imath \xi ^2 \left(5 \xi ^2-1\right) \sinh (2 \pi  \xi ) (\psi ^{(1)}(1-\imath \xi )-\psi ^{(1)}(\imath \xi +1))}{16 \pi  \tau ^4}+\frac{\xi  \left(30 \xi ^2-11\right) \sinh (2 \pi  \xi )}{16 \pi  \tau ^4}\notag\\
	\end{align}
	where $\psi$ is the Digamma function and $\gamma$ is the Euler-Mascheroni constant.

\vspace{1cm}

\begin{center}
\bf{Backreaction: $\textbf{E}\cdot \textbf{B}$}
\end{center}
	
	We now calculate the integral in Eq. \eqref{eqn:integral}. The quantity we are interested in is:
	\begin{equation}
	\langle {\bf E} \cdot {\bf B} \rangle 
	=-\frac{1}{(2\pi)^2\,a^4} \int\,dk\,k^3\frac{\partial}{\partial \tau} \Bigl( |A_+|^2 - |A_-|^2 \Bigr)\,\equiv-\frac{1}{(2\pi)^2\,a^4}\lim_{\Lambda\to\infty}\mathcal{J}(\xi,\tau,\,\Lambda)\,,
	\end{equation}
	where, obviously, $\mathcal{J}$ has not to be confused with the one of the previous Sections and is given by
	\begin{align}
	\mathcal{J}(\xi,\tau,\,\Lambda)=&-\frac{1}{2 \tau}\int_{0}^{\Lambda a}\,dk\,k^2 e^{\pi\xi}\Bigl[W_{\imath \xi,\frac{1}{2}}(-2 \imath k \tau)W_{-\imath \xi+1\,,\frac{1}{2}}(2 \imath k \tau)+ W_{-\imath \xi,\frac{1}{2}}(2 \imath k \tau)W_{\imath \xi+1\,,\frac{1}{2}}(-2 \imath k \tau)  \notag\\
	 &+W_{-\imath \xi,\frac{1}{2}}(-2 \imath k \tau)W_{\imath \xi+1\,,\frac{1}{2}}(2 \imath k \tau)+ W_{\imath \xi,\frac{1}{2}}(2 \imath k \tau)W_{-\imath \xi+1\,,\frac{1}{2}}(-2 \imath k \tau)\Bigr]\,,
	\end{align}
	where, again, we put the IR cutoff to $0$.
	As in the previous section we can use the Mellin-Barnes representation of the Whittaker functions Eq. \eqref{mellin} to write:
	\begin{align}
	\label{iback2}
	&\mathcal{J}(\xi,\tau,\,\Lambda)=-\frac{\xi\sinh^2(\pi\xi)}{2\pi^2\tau}\int_{\mathcal{C}_s}\,\frac{ds}{2\pi\imath}\,\int_{\mathcal{C}_t}\,\frac{dt}{2\pi\imath}\,\frac{\Lambda^{3+s+t}}{3+s+t}(2\imath\tau)^{s+t}\Gamma(-s)\Gamma(1-s)\Gamma(-t)\Gamma(1-t)\\
	&\times\Biggl\{ (\imath+\xi) \left(e^{\imath\pi(s+\imath\xi)}-e^{\imath\pi(t-\imath\xi)}\right) \Gamma(s+\imath\xi-1)\Gamma(t-\imath\xi)+(-\imath+\xi) \left(e^{\imath\pi(t+\imath\xi)}-e^{\imath\pi(s-\imath\xi)}\right)   \Gamma(s-\imath\xi-1)\Gamma(t+\imath\xi) \Biggr\}\,,\notag
	\end{align}
	converging for $\Re{(s+t)}>-3$.
	We only give the final result here since the integral can be carried out as previously explained:
	
	\begin{align}
	\mathcal{J}(\xi,\tau,\,\Lambda)=&\,\frac{\Lambda ^2 \xi }{2 H ^2 \tau ^4}+\frac{3 \xi  \left(5 \xi ^2-1\right) \log (2 \Lambda/H)}{2 \tau ^4}+\frac{3 \gamma  \xi  \left(5 \xi ^2-1\right)}{2 \tau ^4}+\frac{22 \xi -47 \xi ^3}{4 \tau
		^4}+\frac{\left(30 \xi ^2-11\right) \sinh (2 \pi  \xi )}{8 \pi  \tau ^4}\notag\\
	&-\frac{3 \xi  \left(5 \xi ^2-1\right)   \left(H_{-\imath \xi }+H_{\imath \xi }\right)}{4  \tau ^4}+\imath\frac{3 \xi  \left(5 \xi ^2-1\right)  \sinh (2 \pi  \xi ) (\psi ^{(1)}(1-\imath \xi )-\psi ^{(1)}(\imath \xi +1))}{8 \pi  \tau ^4},\notag\\
	\end{align}
	where $H_{x}=\psi(x+1)+\gamma$ is the Harmonic number of order $x$ and $\psi^{(1)}(x)=d\psi(x)/dx$.
	
\end{widetext}

%%%%%%%%%%%%%%%%%%%%%%%%%%%%%%%%%%%%%%%%%%%%

\end{document}